# TorchQC - A framework for efficiently integrating machine and deep learning methods in quantum dynamics and control

Dimitris Koutromanos[a,b], Dionisis Stefanatos[a,c], Emmanuel Paspalakis[a,d]

[a] *Materials Science Department, School of Natural Sciences, University of Patras, 26504 Patras, Greece*

[b] *dkoutromanos@upatras.gr*

[c] *dionisis@post.harvard.edu*

[d] *dkoutromanos@upatras.gr*

**Abstract**

Machine learning has been revolutionizing our world over the last few years and is also increasingly exploited in several areas of physics, including quantum dynamics and control. The need for a framework that brings together machine learning models and quantum simulation methods has been quite high within the quantum control field, with the ultimate goal of exploiting these powerful computational methods for the efficient implementation of modern quantum technologies. The existing frameworks for quantum system simulations, such as QuTip and QuantumOptics.jl, even though they are very successful in simulating quantum dynamics, cannot be easily incorporated into the platforms used for the development of machine learning models, like for example PyTorch. The TorchQC framework introduced in the present work comes exactly to fill this gap. It is a new library written entirely in Python and based on the PyTorch deep learning library. PyTorch and other deep learning frameworks are based on tensors, a structure that is also used in quantum mechanics. This is the common ground that TorchQC utilizes to combine quantum physics simulations and deep learning models. TorchQC exploits PyTorch and its tensor mechanism to represent quantum states and operators as tensors, while it also incorporates all the tools needed to simulate quantum system dynamics. All necessary operations are internal in the PyTorch library, thus TorchQC programs can be executed in GPUs, substantially reducing the simulation time. We believe that the proposed TorchQC library has the potential to accelerate the development of deep learning models directly incorporating quantum simulations, enabling the easier integration of these powerful techniques in modern quantum technologies.

---

## PROGRAM SUMMARY/NEW VERSION PROGRAM SUMMARY

*Program Title:* TorchQC
*Repository link:* https://github.com/qoptics-qtech/torchqc.git
*Licensing provisions:* MIT
*Programming language:* Python
*External libraries:* PyTorch, NumPy, Matplotlib
*Nature of problem:* Control of quantum systems is crucial for developing efficient algorithms that process quantum information and drive quantum systems to desired states. Machine Learning (ML) methods such as Deep Learning, Deep Reinforcement Learning, and basic or advanced optimization methods are increasingly used in Quantum Control by improving existing methodologies or even adding new ones. Simulation of quantum dynamics and Machine Learning methods are not available in a single framework, and one has to combine different libraries to develop a new method. Creating a custom algorithm by combining different libraries or frameworks is not always easy to do, and it requires a substantial amount of time and effort.
*Solution method:* We propose a unified framework, called TorchQC, which exploits the tensor engine of the PyTorch library to create both quantum simulation methods and to easily embody ML methods in quantum dynamics and control. We believe that this framework would help users to easily implement ML methods ready to be applied in quantum systems. TorchQC provides all the necessary tools for simulating closed and open quantum systems, while it paves the way to incorporate ML methods into quantum control problems. Without the need to develop numerical simulation methods, the users will be able to produce quantum simulations from the framework's ready routines and directly use the simulated data in ML and optimization techniques.

## 1. Introduction

The controlled behavior of the dynamics of closed and open quantum systems is an indispensable part of a complete framework for quantum simulations and quantum control problems [1, 2, 3]. Analytical calculations and obtained solutions and are not always available for quantum dynamics; therefore, numerical solutions are necessary for the study of the dynamics of many realistic quantum systems [4, 5, 6, 7, 8, 9, 10, 11, 12]. Machine Learning (ML) and Deep Learning (DL) [13] methods have already begun to revolutionize the simulation of quantum systems and offer novel strategies for the control of quantum dynamics [14, 15, 16, 17, 18, 19, 20, 21, 22, 23, 24, 25, 26, 27, 28, 29].

There are already two very good frameworks for simulations of quantum system dynamics, namely QuTip [30] and QuantumOptics.jl [31]. Both libraries are very good and successful in simulating both closed and open quantum systems, and are rather popular among the quantum technologies community. There are also some libraries that utilize quantum optimal control methodologies, like Krotov's method [7, 9], GRadient Ascent Pulse Engineering (GRAPE) and Chopped RAndom Basis (CRAB) methods [9, 32], and programs of optimal control theory that use a combination of QuTiP with numerical optimization [8], written in Python and Julia programming languages. There are also other software packages that offer quantum optimal control routines capable to simulate in GPU the controlled dynamics of multiqubit systems [10] and in CPU and GPU the controlled dynamics of photo-excited chemical systems [11, 12]. On the other hand, PyTorch [33] is one of the best frameworks for creating deep learning models. Many researchers and companies use PyTorch to develop deep learning solutions. Incorporating libraries, such as QuTip, in PyTorch is not always the easiest case or the best choice, since models with external dependencies are not always easy to maintain. The proposed TorchQC framework comes to fill exactly this gap. It utilizes PyTorch and its tensor mechanism to represent quantum states and operators as tensors, while it also incorporates all the tools needed to simulate quantum system dynamics. It is a solution that comes with low cost since all necessary operations are internal in the PyTorch library. We believe that this library will accelerate the development of DL models directly incorporating quantum simulations into the picture, without breaking any gradient tracking. This is very important if one wants to create differentiable models that can be trained easily and effectively. The library can be used for quantum control and quantum optics [34] problems, mainly using classical algorithms for simulating quantum dynamics, with analytical solutions when possible and numerical solutions when needed, potentially combined with ML models and algorithms that can be implemented in PyTorch.

We note that there is another library that is also based on the PyTorch framework and is mainly used for quantum ML with quantum algorithms, called Torchquantum [35]. This library simulates quantum computations on classical hardware using PyTorch. It is mainly used by researchers for quantum algorithms design. Our proposed library, TorchQC, uses classical algorithms to simulate quantum dynamics and is perfectly suitable for using classical machine learning algorithms and for implementing ML methodologies in quantum control.

## 2. Quantum States

The Hilbert space $\mathcal{H}$ is a normed vector space over $\mathbb{C}$ equipped with an inner product $\langle , \rangle$ that defines the norm as the square root of the self-inner product for each element of the vector space. Each element vector can be represented as a linear combination of basis vectors for a chosen basis in the Hilbert space $\{|\psi_n\rangle\}$:

$$|\psi\rangle = \sum_n c_n |\psi_n\rangle \, , \, c_n \in \mathbb{C} \text{ and } |\psi_n\rangle \in \mathcal{H} \tag{1}$$

This can also be represented as a column vector with entries the complex numbers $c_n$:

$$|\psi\rangle = \begin{pmatrix} c_1 \\ c_2 \\ \dots \\ c_n \end{pmatrix}. \tag{2}$$

Vectors can be regarded as special cases of tensors. A vector can be represented as a rank-1 tensor, which means that only one set of indices is needed ($T_i$) for their definition. Quantum states are mathematically represented in two ways, either as a normalized vector $|\psi\rangle$ in the Hilbert space $\mathcal{H}$ that is associated with the quantum system, or as a positive semi-definite Hermitian operator of trace one acting on the Hilbert space, called density operator or density matrix in finite dimensions ($\rho$). In the TorchQC framework the quantum state can be represented in both ways.

A state vector is represented as a Python class named QuantumState with elements the number of dimensions, the tensor representation, and potentially a list of product dimensions, if the quantum state is a tensor product state.

```
1 from torchqc.states import QuantumState
2
3 qubit_state = QuantumState.basis(2)[0]
4 print(qubit_state.state_tensor)
```

Listing 1: QuantumState definition

```
tensor([[1.+0.j],
        [0.+0.j]], dtype=torch.complex128)
torch.Size([2, 1])
```

In the above code, a qubit basis state has been created, specifically the first basis state $|0\rangle$, using the "basis" method of the QuantumState class. The vectors can participate in all Hilbert space operations, vector addition, scalar multiplication, and inner product. The following code uses this qubit basis to construct an arbitrary quantum state by simply adding the two basis states $|0\rangle + |1\rangle$. The resulting vector is not normalized, so the library gives the option to normalize it directly by calling the normalize method to get

$|\psi\rangle = (|0\rangle + |1\rangle)/\sqrt{2}$. One can also compute inner products between two state vectors as demonstrated in the following code snippet, where is found that $\langle\psi|0\rangle = 1/\sqrt{2}$.

```
from torchqc.states import QuantumState
import numpy as np

basis_states = QuantumState.basis(2)

arbitrary_state = (basis_states[0] + basis_states[1]).
    normalize()

inner_product_value = arbitrary_state.inner_product(
    basis_states[0])
print(inner_product_value)
```

Listing 2: QuantumState inner product

```
tensor([[0.7071+0.j]], dtype=torch.complex128)
```

The methods of the QuantumState class are summarized in Table 1.

| Method | Description |
|---|---|
| inner_product | inner product with another state |
| populations | populations of quantum state |
| normalize | normalizes the vector to represent a state |
| basis | basis quantum states |
| coherent | coherent quantum state (fock basis) |
| dagger | adjoint of the quantum state ($\|\psi\rangle^\dagger = \langle\psi\|$) |

Table 1: Methods of QuantumState class

## 3. Quantum Operators

Operators in a Hilbert space ($\mathcal{H} \rightarrow \mathcal{H}$) act on vectors of the space and have matrix representations. For a Hilbert space of finite dimension $n$, all operators can be represented as square $n \times n$ matrices. They play a crucial role in the mathematical framework of quantum physics, so their addition is crucial for a quantum simulation framework. They are represented as instances of two Python classes named Operator and DynamicOperator. The difference between the two is that an Operator is a static structure that represents a tensor only at one time instance, while a DynamicOperator can represent the matrix corresponding to an operator in many different time steps. A DynamicOperator is quite useful if we want to represent a time-dependent operator such as the Hamiltonian operator in a quantum control problem. It could potentially be used for the time evolution of an operator in the Heisenberg picture.

Operators can be composed (matrix multiplication in this context), added, and also act on space vectors. All operations are implemented within the Operator and DynamicOperator classes, as demonstrated in the following code snippet. Two operators are defined, which are the two Pauli operators $\sigma_x$ and $\sigma_y$. The two operators can be composed or added simply by using the normal Python operations * and $+$. The two Python operations (*,$+$) have been customized inside the Operator class to match the mathematical operations of addition and composition (matrix multiplication in finite dimensions). Thus, instead of implementing a custom addition and composition function for the operators, the user of the library can directly use the Python operations to add and compose quantum operators. The same mechanism applies to the state vectors (QuantumState class) with the vector addition and scalar multiplication.

```
1  from torchqc.operators import Operator, DynamicOperator
2
3  # define operators
4  sigmax = Operator(dims=2, matrix=torch.from_numpy(
5      np.array([[0.j, 1], [1, 0]])))
6  sigmay = Operator(dims=2, matrix=torch.from_numpy(
7      np.array([[0., -1.j], [1.j, 0]])))
8
9  # operators composition
10 composition = sigmax * sigmay
11 print("Composition = ", composition.matrix)
12
13 # operators addition
14 addition = sigmax + sigmay
15 print("Addition = ", addition.matrix)
```

```
16
17 # oprator acts on a state vector
18 zero_state = QuantumState.basis(2)[0]
19 one_state = sigmax.mul(zero_state)
20 print("New state = ", one_state.state_tensor)
```

Listing 3: Operator operations

```
Composition =  tensor([[0.+1.j, 0.+0.j],
        [0.+0.j, 0.-1.j]], dtype=torch.complex128)
Addition =  tensor([[0.+0.j, 1.-1.j],
        [1.+1.j, 0.+0.j]], dtype=torch.complex128)
New state =  tensor([[0.+0.j],
        [1.+0.j]], dtype=torch.complex128)
```

```
1  from torchqc.common_matrices import sigmaX, sigmaZ
2
3  time = np.arange(0, 1, 0.2)
4
5  def Ht(t, args = []):
6      index = int(np.where(time == t.numpy()[0])[0][0])
7
8      if index % 2 == 0:
9          return sigmaX().matrix
10     else:
11         return sigmaZ().matrix
12
13 Ht = DynamicOperator(dims=2, Ht=Ht, time=time)
14 print(Ht.matrix)
```

Listing 4: DynamicOperator instantiation

```
tensor([[[ 0.+0.j,  1.+0.j],
         [ 1.+0.j,  0.+0.j]],
        [[ 1.+0.j,  0.+0.j],
         [ 0.+0.j, -1.+0.j]],
        [[ 0.+0.j,  1.+0.j],
         [ 1.+0.j,  0.+0.j]],
        [[ 1.+0.j,  0.+0.j],
         [ 0.+0.j, -1.+0.j]],
        [[ 0.+0.j,  1.+0.j],
         [ 1.+0.j,  0.+0.j]]], dtype=torch.complex128)
```

DynamicOperator accepts functions, tensors, or operators in the Ht argument, and creates the dynamic operator accordingly. When the input Hamiltonian $H_t$ is a torch.Tensor class from PyTorch, DynamicOperator sets it as its tensor, ignoring any time input. When the Hamiltonian $H_t$ is a static

Operator class instance, DynamicOperator sets it as the Hamiltonian tensor for all time steps. Finally, when the Hamiltonian $H_t$ is a function of time, the tensor is built using this function to retrieve the Hamiltonian tensor for each time step. This gives great flexibility for the user to design and implement the operators for dynamics simulations. One can implement this functionality as in the above Listing 4. Hamiltonian can be defined as a time-dependent function (lines 5-11). To demonstrate the different Hamiltonian matrices that the Hamiltonian function Ht provides in each time step, the function in the Listing 4 returns a Hamiltonian of the form $\sigma_x$ when the time index (sequential number in the time vector) is divided by 2 and $\sigma_z$ otherwise. The function can be provided as an input argument in the construction of the DynamicOperator class. The constructor of the class will automatically detect that the argument is a function and for each time step (argument time) will get the corresponding Hamiltonian matrix time snapshot.

Quantum states can also be represented as density operators ($\mathcal{B}(\mathcal{H}) \rightarrow \mathcal{B}(\mathcal{H})$) or density matrices. The terms density operators and density matrices will be used interchangeably. The density operato, for pure states, is defined as:

$$\rho = |\psi\rangle \langle\psi| , \forall\psi \in \mathcal{H}, \tag{3}$$

or in a matrix representation:

$$\rho = \begin{pmatrix} |a_1|^2 & a_1 a_2^* & \dots \\ a_2 a_1^* & |a_2|^2 & \dots \\ \vdots & \vdots & \ddots \end{pmatrix}. \tag{4}$$

In the TorchQC framework, a quantum state can be converted to a density matrix as follows, where state $|0\rangle$ is converted to the corresponding density matrix:

```
1 from torchqc.states import QuantumState
2 from torchqc.common_functions import get_density_matrix
3
4 zero_state = QuantumState.basis(2)[0]
```

```python
rho = get_density_matrix(zero_state)
print(rho.matrix)
```

Listing 5: Density matrix as Operator

```
tensor([[1.+0.j, 0.+0.j],
        [0.+0.j, 0.+0.j]], dtype=torch.complex128)
```

| **Method** | **Description** |
|---|---|
| mul | acts on a state vector (QuantumState) |
| opmul | composition with another operator |
| dagger | adjoint operator |
| is_hermitian | checks if operator is hermitian |
| is_unitary | checks if operator is unitary |

Table 2: Methods of Operator class

DynamicOperator is an extension of the class of Operator. It inherits all methods from Operator class, with specific implementations for the construct method __init__ and method dagger.

## 4. Tensor product states and operators

For two quantum systems with Hilbert spaces $\mathcal{H}_1$ and $\mathcal{H}_2$, the Hilbert space of the composite quantum system is defined by the tensor product of the two spaces, as $\mathcal{H} = \mathcal{H}_1 \otimes \mathcal{H}_2$. A separable state in the composite system is a vector defined as the tensor product of the two individual quantum states in the constituent Hilbert spaces [3],

$$|\psi\rangle = |\psi_1\rangle \otimes |\psi_2\rangle \,, \text{ where } |\psi_i\rangle \in \mathcal{H}_i, \;\; |\psi\rangle \in \mathcal{H}. \tag{5}$$

The tensor product space $\mathcal{H}$ can be constructed from orthonormal bases of $\mathcal{H}_1$ and $\mathcal{H}_2$. If $\{|e_i\rangle\}$ and $\{|f_i\rangle\}$ are orthonormal bases of $\mathcal{H}_1$ and $\mathcal{H}_2$, respectively, the orthonormal basis of the tensor product space $\mathcal{H}$ is defined as $\{|e_i\rangle \otimes |f_j\rangle\}$, by getting all pair combinations between the two constituent bases. Any state vector in $\mathcal{H}$, including entangled states, can be expressed

in this basis as:

$$|\psi\rangle = \sum_{i,j} a_{ij} |e_i\rangle \otimes |f_j\rangle \tag{6}$$

A quantum simulations library deals only with finite-dimensional Hilbert spaces, where the tensor product between vectors or operators can be represented by the Kronecker product, as follows:

$$\begin{pmatrix} a_1 \\ a_2 \\ \vdots \\ a_n \end{pmatrix} \otimes \begin{pmatrix} b_1 \\ b_2 \\ \vdots \\ b_m \end{pmatrix} = \begin{pmatrix} a_1 \begin{pmatrix} b_1 \\ b_2 \\ \vdots \\ b_m \end{pmatrix} \\ a_2 \begin{pmatrix} b_1 \\ b_2 \\ \vdots \\ b_m \end{pmatrix} \\ \vdots \\ a_n \begin{pmatrix} b_1 \\ b_2 \\ \vdots \\ b_m \end{pmatrix} \end{pmatrix} \tag{7}$$

The same logic applies for the density matrices and operator matrices in general. The library supports composite states or operators representations. Composite states are just as normal states, with the additional information about the individual smaller systems dimensions, so that partial trace of quantum states can be performed. In the following snippet, one can see how a composite state can be constructed with the *tensor_product_states* method while the *partial_trace* method gives the reduced density matrices from the composite density matrix. The second argument of the method defines which subsystem should be traced out.

```
from torchqc.tensor_product import tensor_product_ops, \
```

```python
        tensor_product_states, partial_trace

basis_states = QuantumState.basis(2)
psi_1 = (basis_states[0] + basis_states[1]).normalize()
psi_2 = (basis_states[0] - basis_states[1]).normalize()

psi = tensor_product_states(psi_1, psi_2)
print("Composite state = ", psi.state_tensor)
print("Product dims = ", psi.product_dims)

rho_1 = get_density_matrix(psi_1)
rho_2 = get_density_matrix(psi_2)

print("Density matrix 1 = ", rho_1.matrix)
print("Density matrix 2 = ", rho_2.matrix)

rho = tensor_product_ops(rho_1, rho_2)
print("Composite density matrix = ", rho.matrix)
print("Product dims = ", rho.product_dims)

print("\nTracing out system 2:")
rho_1 = partial_trace(rho, [1])
print("Density matrix 1 (by partial trace): ", rho_1.matrix)

print("Tracing out system 1:")
rho_2 = partial_trace(rho, [0])
print("Density matrix 2 (by partial trace): ", rho_2.matrix)
```

Listing 6: Tensor product states

```
Composite state =  tensor([[ 0.5000+0.j],
        [-0.5000+0.j],
        [ 0.5000+0.j],
        [-0.5000+0.j]], dtype=torch.complex128)
Product dims =  [2, 2]
Density matrix 1 =  tensor([[0.5000+0.j, 0.5000+0.j],
        [0.5000+0.j, 0.5000+0.j]], dtype=torch.complex128)
Density matrix 2 =  tensor([[ 0.5000+0.j, -0.5000+0.j],
        [-0.5000+0.j,  0.5000-0.j]], dtype=torch.complex128)
Composite density matrix =  tensor([[ 0.2500+0.j, -0.2500+0.j,  0.2500+0.j, -0.2500+0.j],
        [-0.2500+0.j,  0.2500+0.j, -0.2500+0.j,  0.2500+0.j],
        [ 0.2500+0.j, -0.2500+0.j,  0.2500+0.j, -0.2500+0.j],
        [-0.2500+0.j,  0.2500+0.j, -0.2500+0.j,  0.2500+0.j]],
       dtype=torch.complex128)
Product dims =  [2, 2]
Tracing out system 2:
Density matrix 1 (by partial trace):  tensor([[0.5000+0.j, 0.5000+0.j],
        [0.5000+0.j, 0.5000+0.j]], dtype=torch.complex128)
```

```
Tracing out system 1:
Density matrix 2 (by partial trace):  tensor([[ 0.5000+0.j, -0.5000+0.j],
        [-0.5000+0.j,  0.5000+0.j]], dtype=torch.complex128)
```

| Method | Description |
|---|---|
| tensor_product_states | tensor product of QuantumState instances |
| tensor_product_ops | tensor product of Operator instances |
| partial_trace | partial trace of the given composite state |

Table 3: Tensor product operations

| Matrix | Description |
|---|---|
| sigmaX | Pauli matrix $\sigma_x$ |
| sigmaY | Pauli matrix $\sigma_y$ |
| sigmaY | Pauli matrix $\sigma_z$ |
| sigmaPlus | raising operator $\sigma_+$ (dims=2) |
| sigmaMinus | lowering operator $\sigma_-$ (dims=2) |
| annihilation | annihilation operator $a$ |
| creation | creation operator $a^\dagger$ |
| eye | identity operator |
| displacement_operator | displacement operator |

Table 4: Common matrices methods

## 5. Quantum Dynamics

The quantum system dynamics is determined by the Hamiltonian operator [1, 3]. This operator is involved in the fundamental equations that describe the time evolution of a quantum state, in the Schrödinger equation and in the Lindblad-von Neumann equation. The Hamiltonian operator can also be time-dependent, a usual case in quantum control problems, where the Hamiltonian also incorporates the interactions of the system with external fields.

| Method | Description |
|---|---|
| TDSE | solves the Time Dependent Schrödinger Equation analytically |
| TDSE_numeric | solves the Time Dependent Schrödinger Equation numerically |
| lindblad_equation | solves lindblad master equation numerically |
| lindblad_equation_FLS | solves Fock-Liouville space dynamics |

Table 5: Quantum dynamics simulation methods

### *5.1. Schrödinger equation*

Let $|\psi(t)\rangle$ describe the state of a quantum system at time $t$. The state evolution in time, for closed quantum systems (no dissipation or losses), is given by the Schrödinger equation:

$$i\hbar\frac{d}{dt}|\psi(t)\rangle = H(t)|\psi(t)\rangle \,. \tag{8}$$

The solution to this equation at time $t = t_1$ is formally expressed as

$$|\psi(t_1)\rangle = U(t_1, t_0)|\psi(t_0)\rangle \,, \tag{9}$$

where

$$U(t, t_0) = \mathcal{T}e^{-i\hbar\int_{t_0}^{t} H(t')dt'} \tag{10}$$

is the unitary propagator operator and $\mathcal{T}$ denotes the time ordering operator. This propagator satisfies the operator Schrödinger equation:

$$i\hbar\frac{\partial}{\partial t}U(t, t_0) = H(t)U(t, t_0). \tag{11}$$

The Time Dependent Schrödinger Equation (TDSE method) is implemented in the TorchQC library. Its solution can be both numerical and analytical (9). In the analytical case, during each time step the Hamiltonian is considered to be constant. In the following code, TDSE method is used to simulate the Rabi oscillations in a qubit:

```python
import numpy as np
import torch

from torchqc.states import QuantumState
from torchqc.common_functions import get_density_matrix,
    expect_val_dm
from torchqc.operators import DynamicOperator
from torchqc.common_matrices import sigmaY, sigmaX
from torchqc.dynamics import TDSE
import matplotlib.pyplot as plt

# define the initial quantum state
n = 2
basis_states = QuantumState.basis(n)
initial_state = basis_states[0]

# Simulate the dynamics with the TDSE
T = 10
Dt = 0.1
time = np.arange(0, T + Dt, Dt, dtype = np.float32)

hamiltonian = DynamicOperator(dims=n, Ht=sigmaX(), time=time)
states = TDSE(initial_state, hamiltonian, time, Dt)

# print the rabi oscillations
populations = np.array([state.populations() for state in
    states])

fig, ax = plt.subplots()
ax.plot(time, populations[:,0], label = "P1")
ax.plot(time, populations[:,1], label = "P2")
ax.legend()
fig.show()
```

Listing 7: TDSE dynamics

In the snippet above, it is easy to see that the requirements for the simulation is the Hamiltonian operator, defined as a DynamicOperator, the initial state of the system as a state vector, the array of the times steps and the time step interval. Fig. 1 is the output of the above program, demonstrating the Rabi oscillations experienced by the qubit system, during which the two states exchange populations.

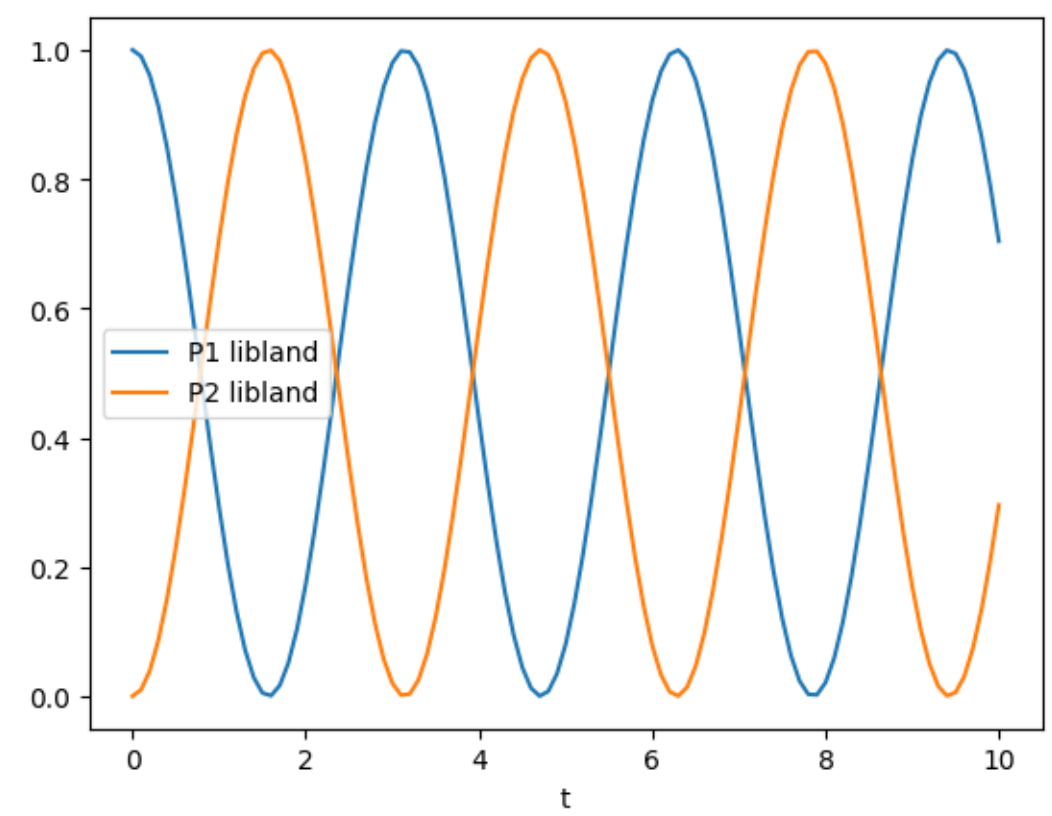

Figure 1: Qubit Rabi oscillations, solution by TDSE method.

### 5.2. Lindblad-von Neumann equation

The other fundamental equation of quantum dynamics is the Lindblad-von Neumann equation, which describes the evolution of the density matrix in both closed and open quantum systems. For closed systems the equation reads:

$$i\hbar\frac{d}{dt}\rho = [H(t), \rho], \tag{12}$$

while for open systems:

$$\frac{d}{dt}\rho = -\frac{i}{\hbar}[H(t), \rho] + \sum_i \gamma_i (L_i \rho L_i^\dagger + \frac{1}{2}\{L_i^\dagger L_i, \rho\}), \tag{13}$$

where $L_i$ are the jump operators that describe the dissipative behavior of the system and $\{\cdot,\cdot\}$ is the anticommutator billinear operator, defined as follows:

$$\begin{aligned} \{\cdot,\cdot\} : \mathcal{B}(\mathcal{H}) \times \mathcal{B}(\mathcal{H}) &\to \mathcal{B}(\mathcal{H}) \\ (\rho, \sigma) &\mapsto \{\rho, \sigma\} = \rho\sigma + \sigma\rho. \end{aligned} \tag{14}$$

The Lindblad equation is solved numerically within the TorchQC library using Runge–Kutta methods (i.e. rk45, rk4) or Adams (linear multistep) methods. The main numerical method used is the Runge-Kutta method of order 5(4) (Runge–Kutta–Fehlberg method rk45). Runge–Kutta methods are numerical methods that solve ordinary differential equations (ODEs) [36]. They are iterated methods composed of several stages. A weighted average of the solution function is computed at nearby points to give the solution at the next time step. The simplest method for solving ODEs is the Euler's method. RK45 method, as can be derived from its name, is a method of order $O(h^4)$ with an error estimator of order $O(h^5)$. Multistep methods aim to improve efficiency by retaining and utilizing information from prior steps instead of discarding it. As a result, these methods rely on multiple previous points and their corresponding derivative values. Specifically, in linear multistep methods, a linear combination of these earlier points and derivatives is employed.

In the code snippet bellow, the dynamics of a single qubit with Hamiltonian

$$H = \frac{\Delta}{2}\sigma_x \tag{15}$$

is simulated using the Lindblad master equation with collapse operator $L = \sigma_z$ and rate $\gamma = 0.25$, and the expectation value of operator $\sigma_z$ is plotted in Fig. 2. The numerical results are compared with the analytical solution given by:

$$\langle \sigma_z \rangle = e^{-\gamma t} \cos(2\pi t), \tag{16}$$

to evaluate the performance of the library.

One has to keep in mind that since this equation is solved numerically, the time step used should be tuned to acquire convergence. Big time steps may lead to non-steady solutions. Simulating Markovian dissipative quantum dynamics is straightforward within the library. There are ready-to-use methods such as *lindblad_ equation* and *expect_ val_ dm* that produce the simulated density matrices for all the time steps from $t_0$ to $T$ and the expectation values of quantum operators ($S_z$ in the following code), respectively.

```
1 import numpy as np
2 import torch
3 import matplotlib.pyplot as plt
```

```python
4
5 from torchqc.states import QuantumState
6 from torchqc.common_functions import get_density_matrix,
    expect_val_dm
7 from torchqc.operators import DynamicOperator
8 from torchqc.common_matrices import sigmaZ, sigmaX
9 from torchqc.dynamics import lindblad_equation
10
11 # define the initial condition
12 T = 5
13 Dt = 0.02
14 time = np.arange(0, T, Dt, dtype = np.float32)
15 D = 2 * np.pi
16
17 hamiltonian = DynamicOperator(dims=n, Ht=(D / 2) * sigmaX(),
    time=time)
18
19 # simulate the dynamics
20 rho = get_density_matrix(state)
21 jump_operator = sigmaZ()
22 gamma = 0.25
23 time_tensor, lindblad_states = lindblad_equation(rho,
    hamiltonian, time,
24             Dt, [jump_operator], [gamma])
25
26 # find the expection values of Sz
27 Sz_analytic = torch.cos(2 * np.pi * time_tensor) * \
28     torch.exp(-time_tensor * gamma)
29 Sz = expect_val_dm(lindblad_states, sigmaZ())
30
31 # plot the expectation values
32 fig, ax = plt.subplots()
33
34 plt.scatter(time, Sz, c="r", marker="x", label=r"$<S_z>$
    TorchQC")
35 plt.plot(time, $\sigma_{z_analytic}$, label=r"$<S_z>$
    Analytic")
36 ax.legend()
37 fig.show()
```

Listing 8: von Neumann-Lindblad equation dynamics

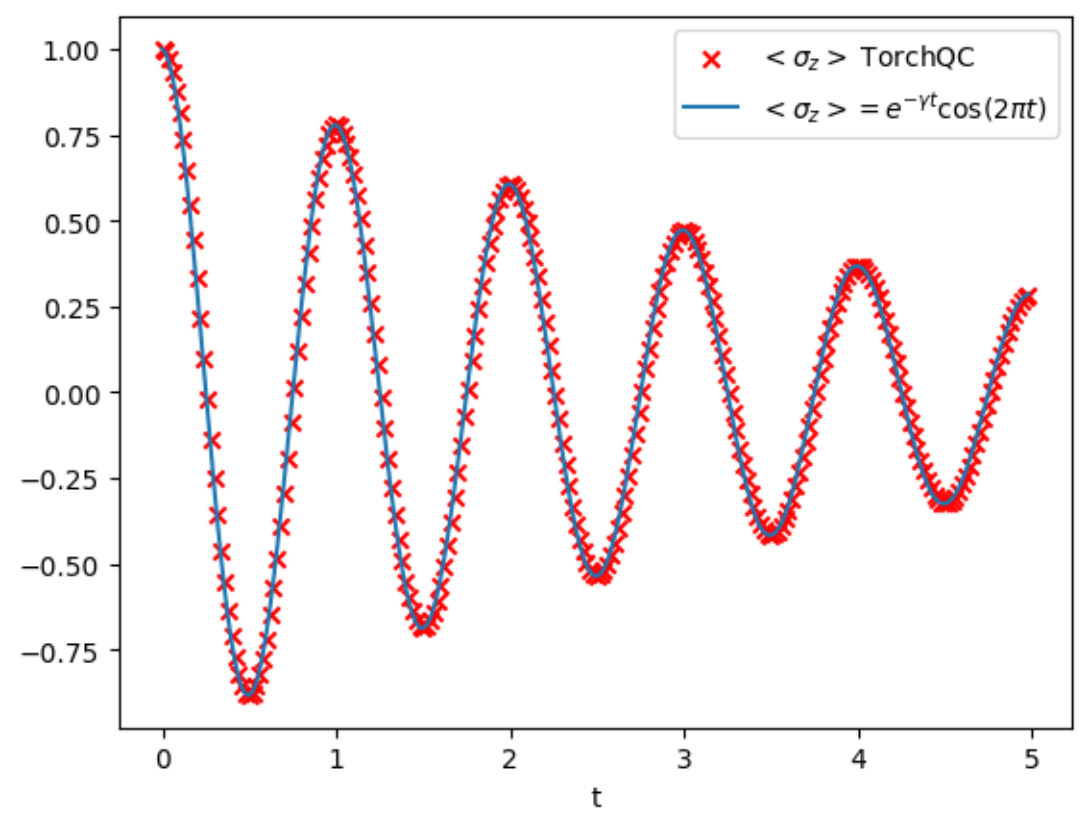

Figure 2: Expectation $\langle \sigma_z \rangle$ (crosses) as obtained by simulating with the TorchQC library the quantum master equation, with jump operator $L = \sigma_z$ and dumping rate $\gamma = 0.25$. Comparison with the analytical solution $\langle \sigma_z \rangle = e^{-\gamma t} \cos(2\pi t)$ (solid line).

*5.3. Fock-Liouville space*

The Fock-Liouville space (FLS) is defined as the tensor product of two Hilbert spaces [37, 38, 39] which, in the context of open quantum systems, are the Hilbert space of the corresponding closed quantum system $\mathcal{H}$ and its dual space $\mathcal{H}^*$, thus $\mathcal{FS} = \mathcal{H} \otimes \mathcal{H}^*$ [38]. In this space, operators are represented as vectors (rank-1 tensors) denoted by double ket symbols, i.e., the density matrix in the FLS is written as $|\rho\rangle\rangle$. For an arbitrary n-level quantum system:

$$|\rho\rangle\rangle = \left(\rho_{11}\ \rho_{12} \ldots \rho_{1n}\ \rho_{21}\ \rho_{22} \ldots \rho_{2n} \ldots \rho_{n1}\ \rho_{n2} \ldots \rho_{nn}\right)^T \tag{17}$$

where $\rho_{ij} \in \mathbb{C}$ are the elements of the density matrix $\rho \in \mathcal{H}$. FLS is itself a Hilbert space equipped with the inner product:

$$\langle\langle A|B\rangle\rangle := Tr(A^\dagger B), \tag{18}$$
$$\text{where } A, B \in \mathcal{B}(\mathcal{H}) \text{ and } |A\rangle\rangle, |B\rangle\rangle \in \text{FLS}$$

If one transitions to the FLS, the quantum dynamics is no longer governed by the Schrödinger equation, but by the special von-Neuman equation which resembles it:

$$\frac{d}{dt}|\rho\rangle\rangle = \mathcal{L}|\rho\rangle\rangle, \tag{19}$$

where $\mathcal{L}$ is a Liouville superoperator that acts on the FLS vectors. Superoperators can be defined using left and right Hilbert-space operators and combinations of them. The left and right superoperators of an operator $O \in \mathcal{H}$ are defined as:

$$L[O]\,|\rho\rangle\rangle = O\rho, \tag{20}$$

$$R[O]\,|\rho\rangle\rangle = \rho O. \tag{21}$$

Using the left and right superoperators, one can also define the Lie bracket operation as a combination of the two:

$$[O,\rho] = O\rho - \rho O = (L[O] - R[O])(\rho), \tag{22}$$

and we write

$$[O,\cdot] = L[O] - R[O]. \tag{23}$$

Combining all the above, the Liouville operator in Eq. (19) can be expressed as:

$$\mathcal{L} = -\frac{i}{\hbar}(L[H] + R[H]) + \sum_i \gamma_i D[F_i], \tag{24}$$

where $D[F_i]$ is the dissipative application of the jump operator $F_i$ defined as follows

$$\begin{aligned} D[F_i](\rho) &= F_i \rho F_i^\dagger - \frac{1}{2}\{F_i^\dagger F, \rho\} \\ &= L[F_i]R[F_i^\dagger](\rho) \\ &- \frac{1}{2}\left[L[F_i^\dagger F_i](\rho) + R[F_i^\dagger F_i](\rho)\right]. \end{aligned} \tag{25}$$

The solution of the FLS equation is given by:

$$|\rho(t)\rangle\rangle = \mathcal{T} e^{\int_0^t \mathcal{L} dt'}\,|\rho(0)\rangle\rangle. \tag{26}$$

The FLS formalism has been implemented in the TorchQC library and the corresponding method gives analytical solutions, similar to the TDSE

method, in contrast to the Lindbladian numerical approach. Analytical solutions mean that during the time step interval, the Lindblad superoperator $\mathcal{L}$ is considered constant. This enables us to solve the differential equation directly and propagate the density matrix elements in time using the analytical solution of the exponential of a constant matrix, as shown in Eq. (26). The disadvantage of the FLS formalism is that the quantum states are mapped into a larger Hilbert space, which can be computationally expensive for high-dimensional quantum systems. Otherwise, FLS is a more preferable approach over the numerical solution of the Lindblad equation, since its accuracy does not depend on the time step discretization. The code snippet demonstrating the use of FLS method for solving the dynamics of a dissipative two-level quantum system can be found in Appendix A. The results are displayed in Figs. 3 and 4.

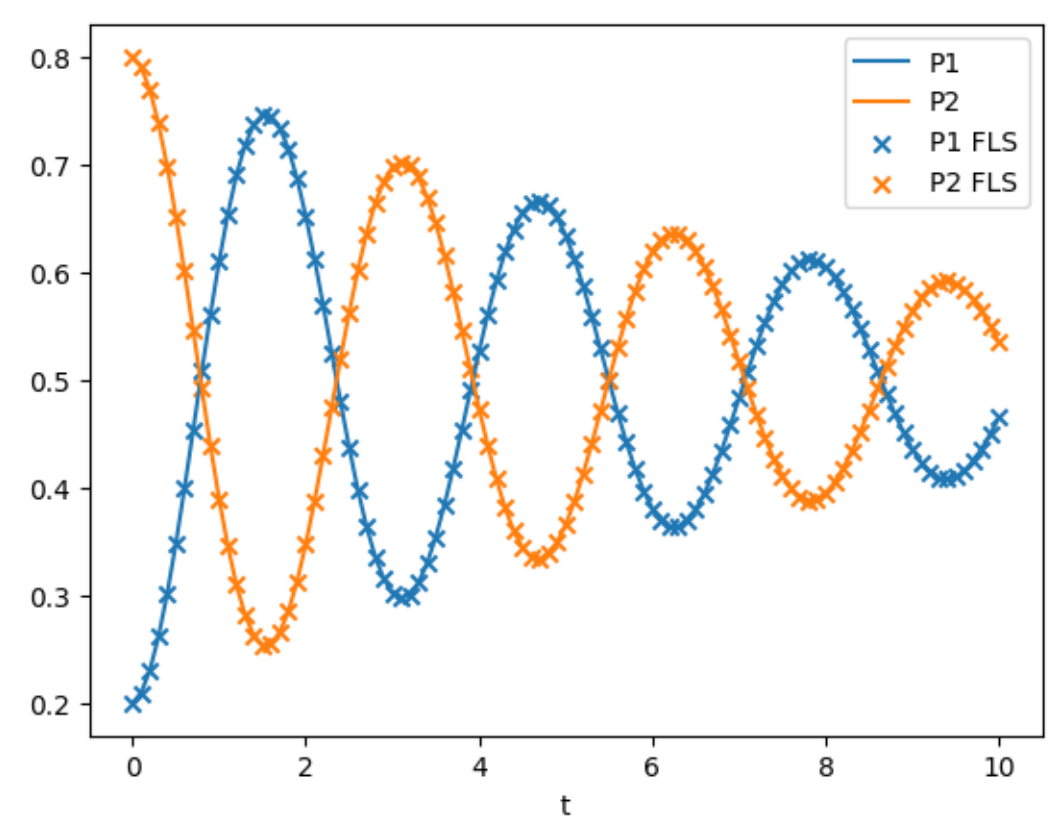

Figure 3: Populations of states 1 and 2 in a dissipative two level system, calculated by solving numerically the Lindblad equation (solid lines) and analytically with the FLS method (crosses).

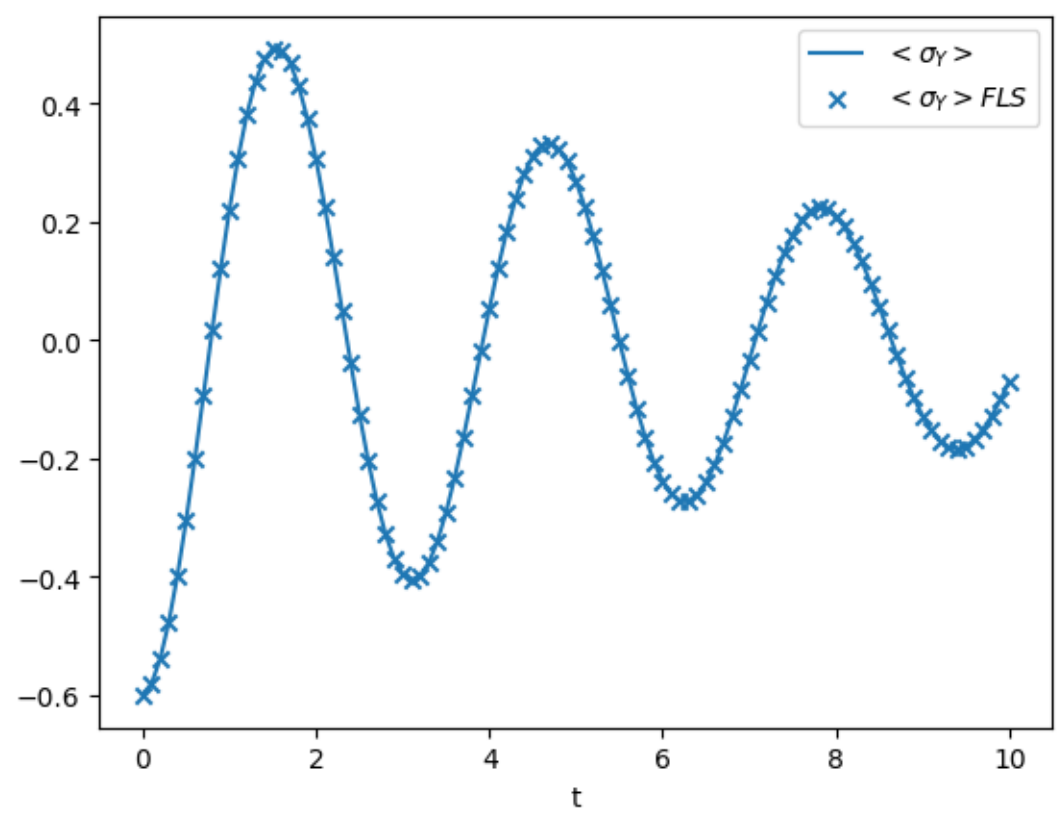

Figure 4: Expectation value $\langle \sigma_y \rangle$ calculated with the numerical Lindblad (solid) and analytical FLS (crosses) methods.

## 6. Quantum Control

Quantum Control is the discipline that studies the interaction of quantum systems with electromagnetic fields and how external fields can be designed to steer these systems to desired target states [1, 3]. Many methods can be used to obtain the appropriate shapes of the fields achieving the desired outcome. These include resonant pulse methods [1], adiabatic methods like rapid adiabatic passage [1, 40] and stimulated Raman adiabatic passage [41], shortcuts to adiabaticity [42] and optimal control [43, 44, 7], which may be utilized to produce pulse shapes solving the various quantum control problems. Over the last years, with the recent advances in machine learning and particularly deep learning and reinforcement learning, new quantum control techniques have emerged, incorporating these powerful methods.

### *6.1. GRAPE with PyTorch Automatic Differentiation*

One of the main quantum optimal control algorithms is GRAPE [32, 9, 45, 46, 6]. The algorithm discretizes time and during each time step interval the applied controls are constant. The propagator associated with the $k$-th step is

$$U(t_k, t_{k-1}) = e^{-iH(t_k - t_{k-1})}, \tag{27}$$

where the Hamiltonian is constant during the time interval $Dt = t_k - t_{k-1}$. The total propagator corresponding to the applied piecewise constant controls is given by the product of the individual propagators for all time steps:

$$U(T,0) = U(t_N, t_{N-1}) \dots U(t_2, t_1) U(t_1, t_0), \tag{28}$$

where $t_0 = 0$ is the initial time, N is the number of time steps and $t_N = T$ is the final time. Each propagator can be parameterized with the control values which construct the Hamiltonian during the corresponding time step, so the product becomes:

$$U(T, 0|\boldsymbol{\theta}) = U(t_N, t_{N-1}|\boldsymbol{\theta_N}) \dots U(t_2, t_1|\boldsymbol{\theta_2}) U(t_1, t_0|\boldsymbol{\theta_1}), \tag{29}$$

where $\boldsymbol{\theta}$ is the vector of all parameters used in the product and $\boldsymbol{\theta_i}$ is the parameter vector used for the $i$-th component. In the GRAPE method, these parameters are optimized using a gradient algorithm to maximize the fidelity between the final and the target states.

GRAPE can be implemented with the TorchQC proposed platform. This is demonstrated using a simple qubit system, described by the following Hamiltonian under the electric dipole and rotating wave approximations

$$H(t) = \frac{\hbar}{2} \begin{bmatrix} \Delta(t) & \Omega(t) \\ \Omega(t) & -\Delta(t) \end{bmatrix} = \hbar \frac{\Delta(t)}{2} \sigma_z + \hbar \frac{\Omega(t)}{2} \sigma_x. \tag{30}$$

The two piecewise constant control functions are the Rabi frequency $\Omega(t)$ and the detuning $\Delta(t)$, which are considered to be bounded. Their constant values in each time step define the corresponding optimization parameters. The optimization goal is the complete population inversion, starting from state 1, thus the objective function is to maximize the population of state 2 or minimize the infidelity between state 2 and the final state achieved with the dynamics.

The code that implements the GRAPE algorithm for this control problem with the help of the TorchQC library can be found in Appendix A. One can call ready functions from the library to define quantum states (Sec. 1) and quantum operators (Sec. 4) such as the Pauli matrices. In the code snippet, the unitary gate corresponding to each propagator in Eq. (29) is defined as a module of the PyTorch package, while the product of these unitary gates is constructed as another PyTorch module. In this way, the parameters of each gate can be trained directly by the automatic differentiation mechanism of PyTorch, while trying to maximize the fidelity between the final state and the target state. In the training process, the stochastic gradient descent optimization algorithm Adam has been employed. The loss function also punishes very large control values, and in combination with the imposed control bounds, it is ensured that the controls remain low enough for feasible experimental setups, while achieving a very high final fidelity value. The optimized control parameters obtained after training are plotted in Fig. 5, while the corresponding evolution of populations is shown in Fig. 6. The final fidelity of the target state is equal to 0.9997. This is a simple demonstration of how one can setup simple machine learning models and implement quantum optimal control algorithms with the help of the library. The integration of quantum problems within the context of PyTorch enables many possibilities offered by this library, as demonstrated with the use of the Adam optimization algorithm which is already implemented in PyTorch.

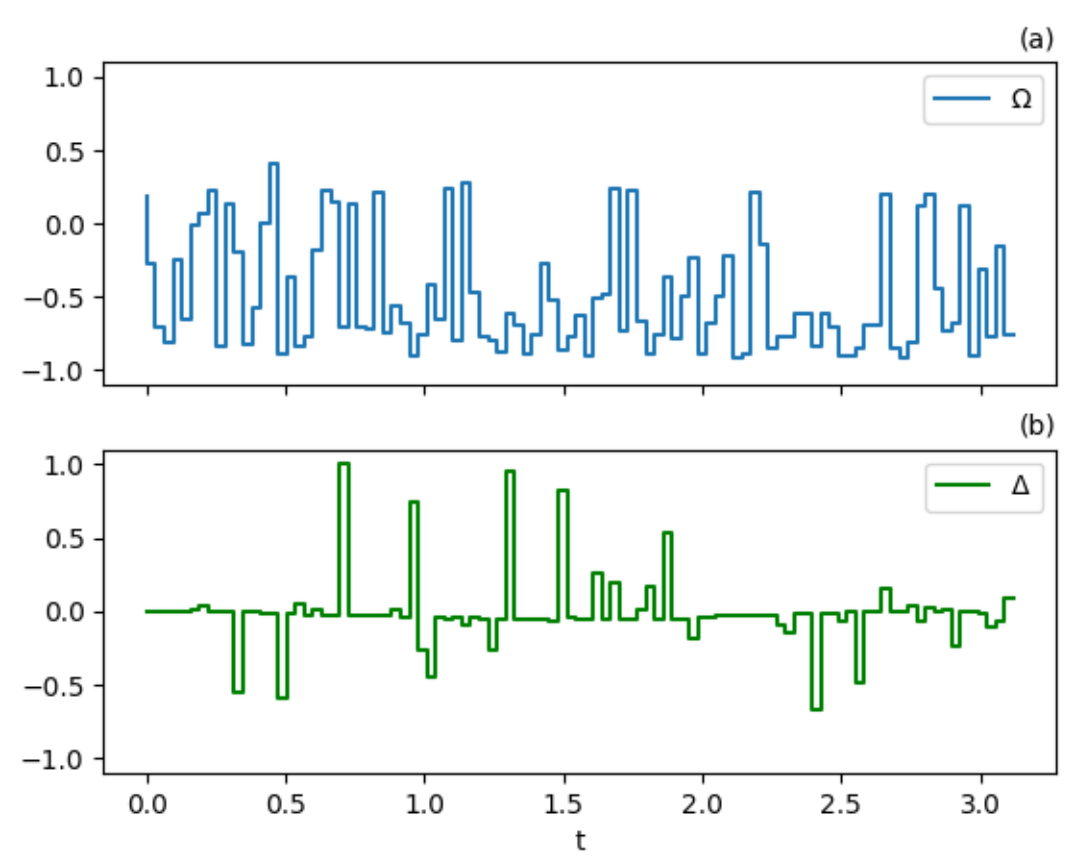

Figure 5: Piecewise constant control functions obtained with GRAPE. (a) Rabi frequency $\Omega(t)$, (b) Detuning $\Delta(t)$.

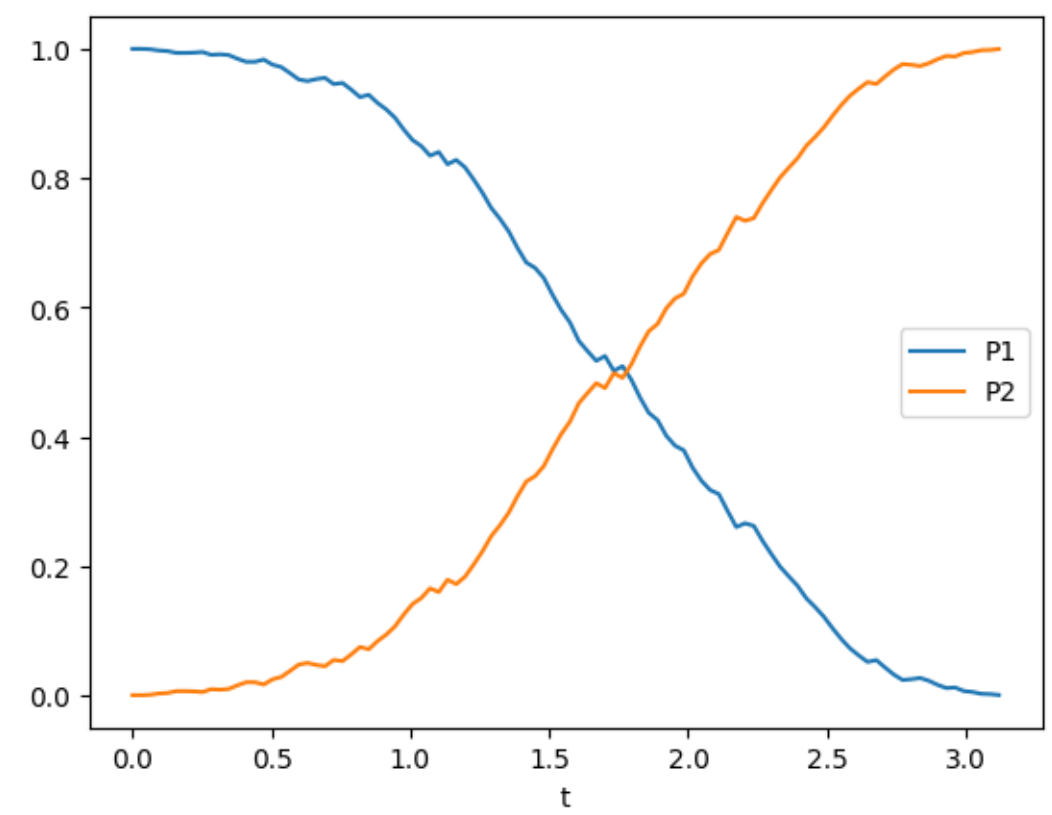

Figure 6: Evolution of populations of qubit states under the GRAPE controls.

### *6.2. Control with Deep Reinforcement Learning*

The qubit population inversion quantum control problem solved previously with the GRAPE algorithm, can also be solved by utilizing Reinforcement Learning (RL) techniques. RL algorithms can solve a problem formalized as a Markov Decision Process (MDP). The quantum system dynamics can be formulated as a MDP and several RL methods can be applied to produce a policy, which corresponds to an optimal control function driving the system to the desired state. There are many works utilizing RL methods to enable quantum technologies [14, 15, 16, 17, 18, 19, 20, 21, 22, 23, 24, 25, 26, 27].The crucial step is to design the environment in which the RL agent operates and interacts with. This environment should receive the action of the agent, which in the quantum control case are the control function values, and produce the reward of that action and the new state of the agent-environment system. Figure 7 shows schematically how the agent and the environment interact at a single time step $t$. RL methods make use of the MDP variables, such as the state, reward, and actions, to construct algorithms that can learn from these interactions. After sufficient training, the agent should be able to produce an optimal policy, in the form of control inputs, which can drive the system optimally to the target.

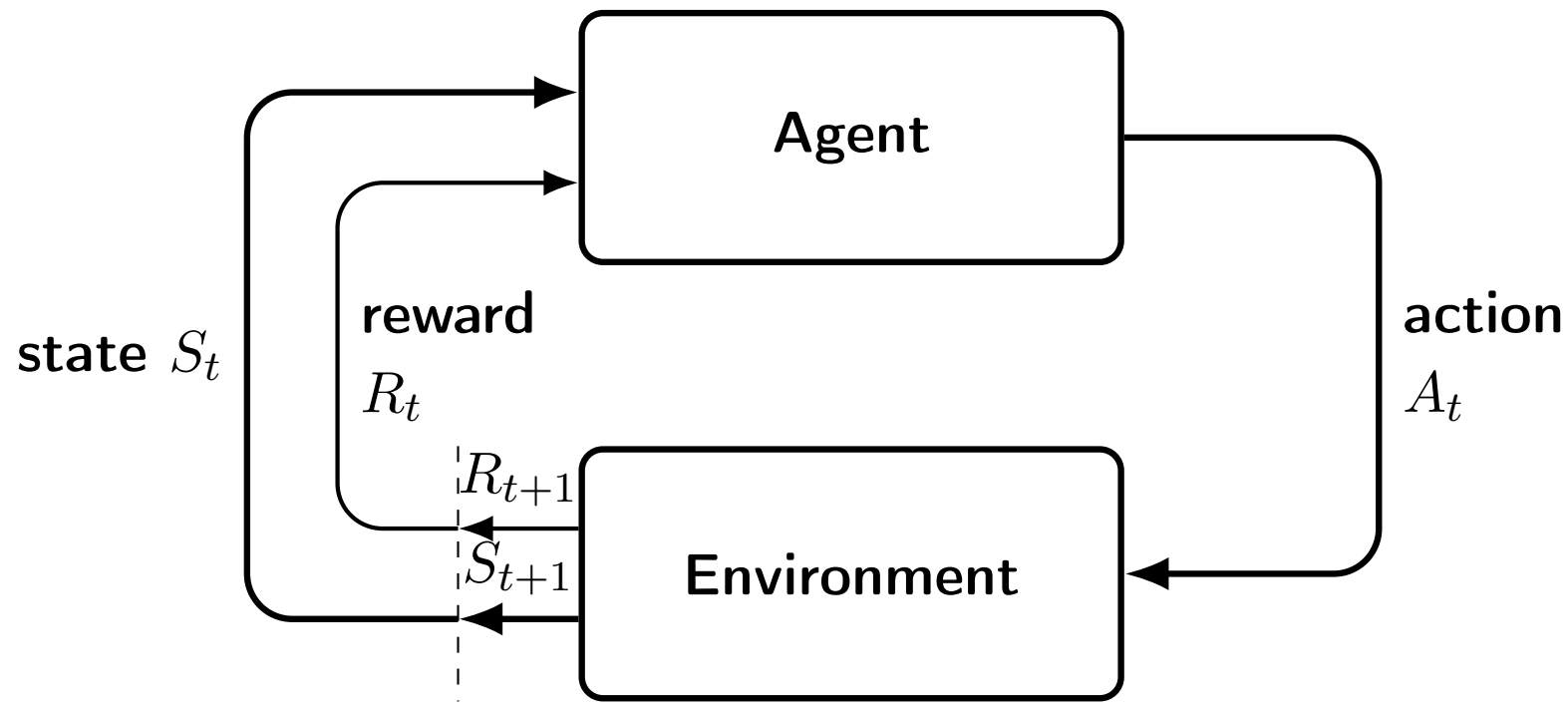

Figure 7: Markov decision process and agent-environment interplay.

For the environment, the qubit system (30) will be used again. For this example, the system is taken to be on resonance, by setting the detuning $\Delta = 0$ for all times, while the only control function used is the Rabi frequency, which is considered to be bounded. The goal is to find a time-dependent Rabi frequency that transfers the population from the ground state to the excited state. Table 6 describes how the corresponding MDP spaces are defined. The actions space includes the available actions ($A_t = \Omega_t$) that the RL agent can perform, which is just the value of the Rabi frequency at a particular time step, while the state space includes all the possible states ($S_t$) the RL agent can assume during an episode. In this example, the state of the RL agent coincides with the quantum state (density matrix elements $\{\rho_{ij}\}$). The reward space has scalar real-valued elements and it is a metric on how good the RL agent performs. The reward at each time step ($R_t$) is equal to $\sqrt{\mathcal{F}(t)} - 1$, where $\mathcal{F}(t) = \mathrm{Tr}[\rho_{tar}\rho(t)]$ is the fidelity of the current state with the target state. The agent is rewarded if the current fidelity is close to the threshold fidelity, but each time step is punished with a constant -1 value. In this way, the agent, which is constantly trying to increase the rewards, will try to find the optimal path with a smaller number of time steps, eventually identifying time-optimal policies. At the final step of the agent's trajectory, the reward may also be increased by a constant value $c > 0$, if the fidelity threshold has been achieved. If the agent passes the threshold, the term

$cu(\mathcal{F}(t))$, where

$$u(\mathcal{F}) = \begin{cases} 1, & if \mathcal{F} \geq \mathcal{F}_{th} \\ 0, & else \end{cases} . \tag{31}$$

is the unit step function, is activated and increases the reward of the agent by the value $c$.

| Space | Element definition |
|---|---|
| Action space | $A_t = \Omega_t$ |
| State space | $S_t = \rho_{ij}$ where $i, j \in \{1, 2\}$ |
| Reward space | $R_t = \begin{cases} \sqrt{\mathcal{F}(t)} + cu(\mathcal{F}(t)), \\ \text{if episode terminates} \\ \sqrt{\mathcal{F}(t)} - 1, \quad \text{else} \end{cases}$ |

Table 6: MDP spaces definitions

One can use the TorchRL library [47], an open-source RL library for PyTorch, to create the algorithms and the environments in the form of a MDP. It provides a lot of functionalities in this direction and also contains a lot of examples of how to make use of its tools. In the corresponding code in Appendix A, the MDP environment for the qubit system is created and the TDSE method from the TorchQC quantum dynamics library is used to propagate the system forward in time at each time step. After the environment has been created, one has to define an algorithm or utilize an existing one from the RL literature. In this context, the Proximal Policy Optimization (PPO) algorithm [48] is used to produce an optimal policy for the complete population inversion problem. More details on how this problem can be formalized as an MDP and how to utilize different kinds of RL methods can be found in our recent work [24].

The results from the training of the PPO agent, displayed in Figs. 8 and 9, show that the optimal policy for the complete state transfer recovers the well-known $\pi$-pulse. The agent can produce this result, known from quantum control theory, without incorporating any explicit physics model in its architecture, only by using the interaction with the qubit environment and by utilizing the actions, the states, and the rewards of the agent-environment system.

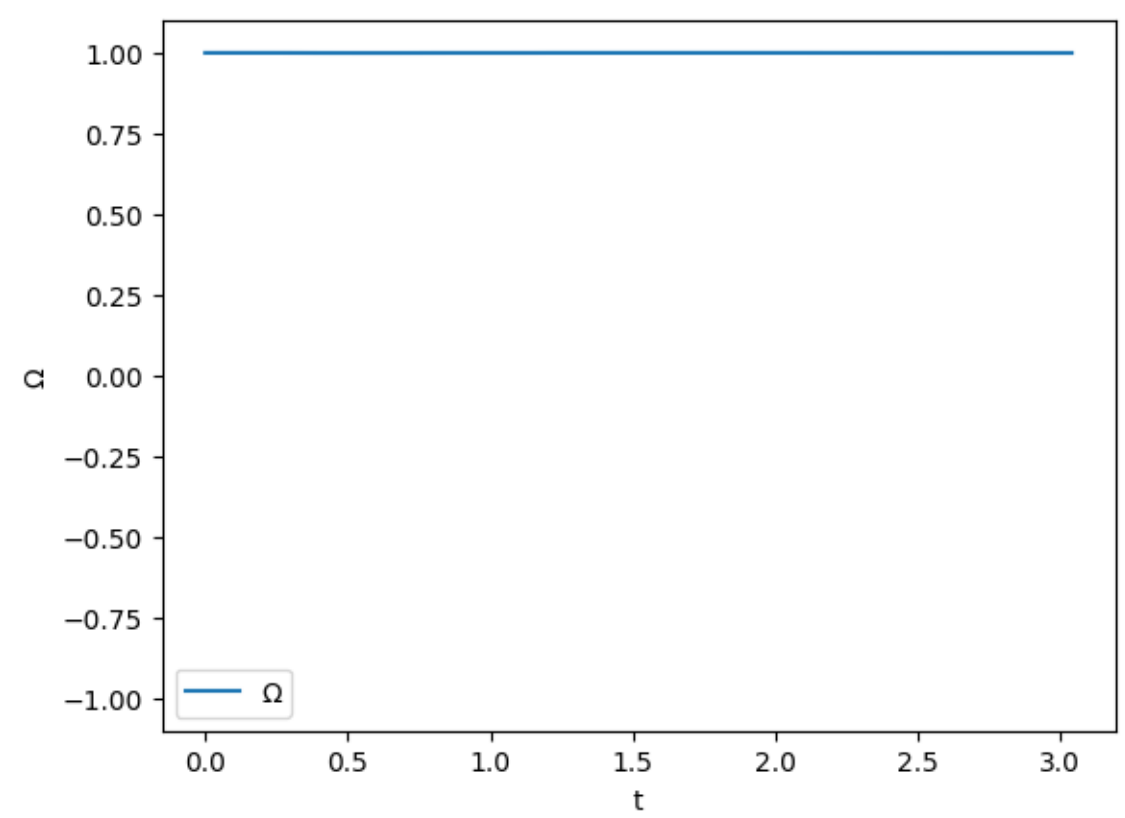

Figure 8: Optimal Rabi frequency obtained with deep reinforcement learning. Note that the system is taken on resonance ($\Delta = 0$).

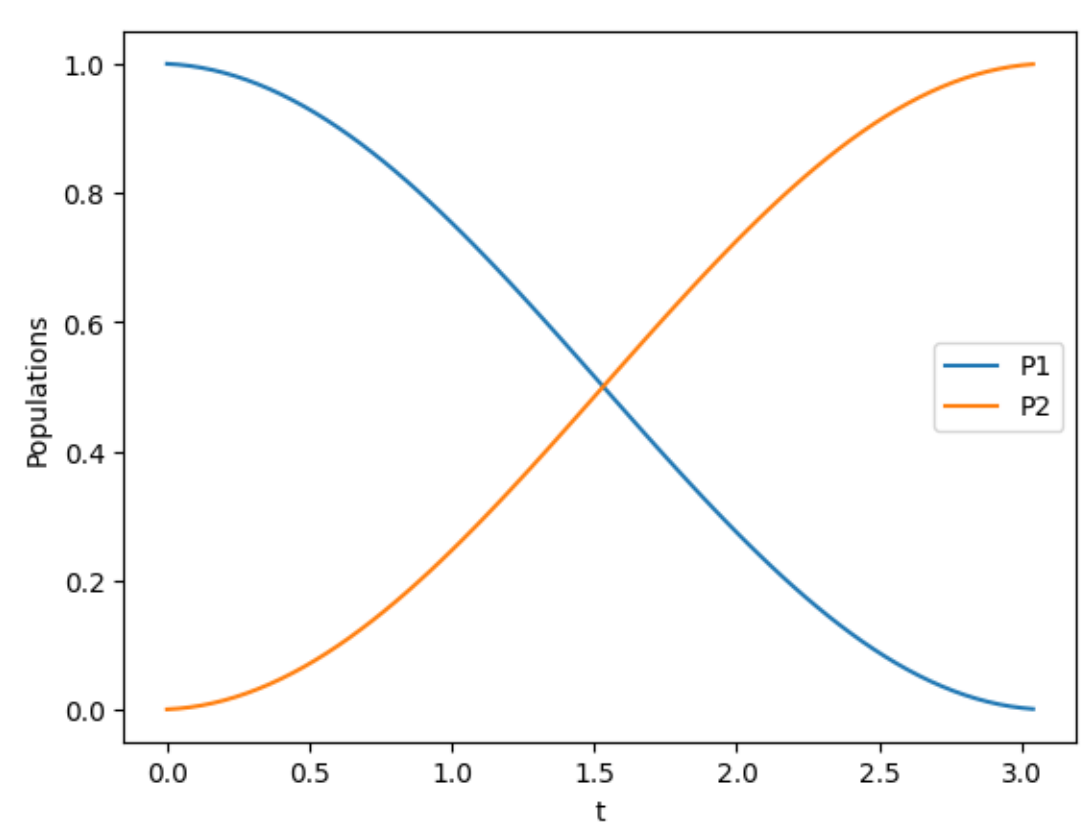

Figure 9: Evolution of populations of qubit states under the control obtained with deep RL.

## 7. Quantum Optics

One of the fundamental systems of Quantum Optics is the Jaynes-Cummings model [49, 50]. It is the fully quantized version of the Rabi model assuming a quantized field expressed in terms of the Fock states [51]. The model describes the interaction and dynamics of a composite system which consists of a qubit interacting with a single-mode quantized field in a cavity. Under the RWA, the system is described by the following Hamiltonian ($\hbar = 1$):

$$H = -\frac{1}{2}\omega_a \sigma_z + \omega_c a^\dagger a + g(\sigma_+ a + \sigma_- a^\dagger) \tag{32}$$

In Appendix A we provide the code simulating the Jaynes-Cummings model with a truncated Fock space for the field. The code produces the expectation value of the population difference of the qubit inside the cavity

$$W(t) = \langle \sigma_z \rangle \,, \tag{33}$$

when starting from a coherent state.

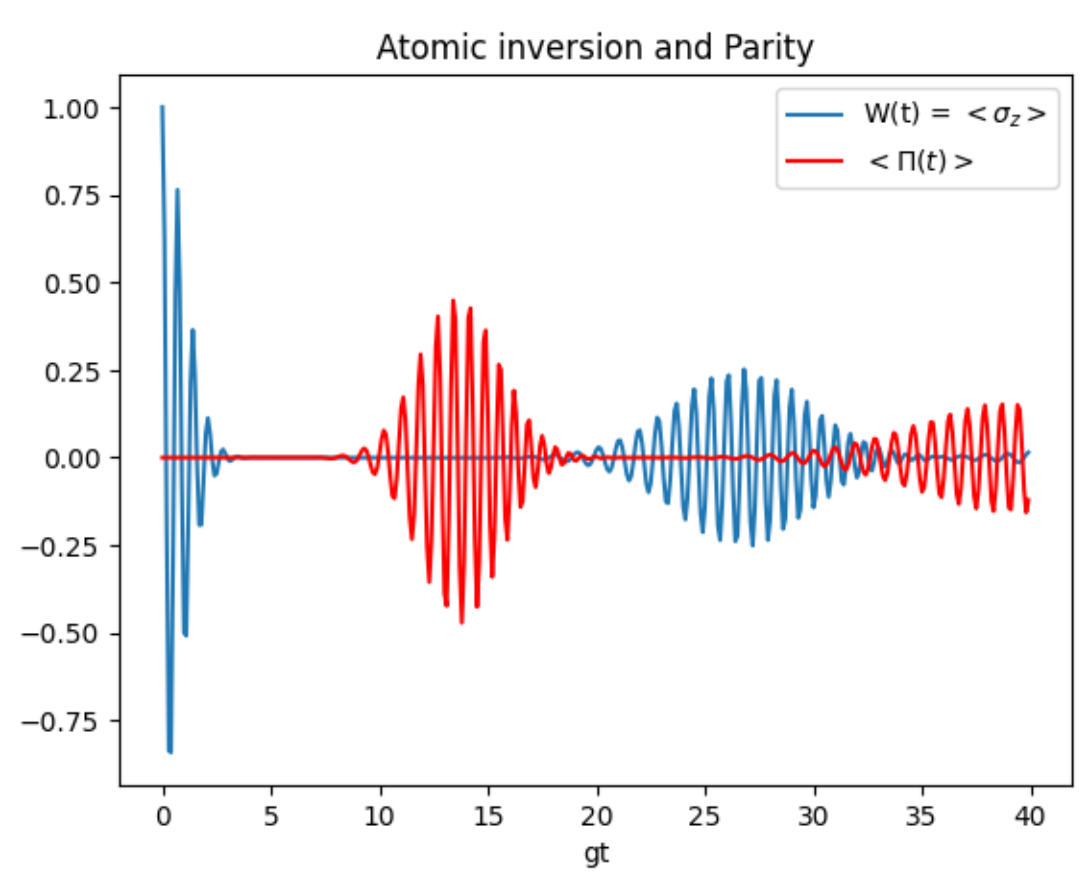

Figure 10: Population difference evolution in Jaynes-Cummings model without losses starting initially with a coherent state.

It is clear for Fig. 10 that the atomic inversion $W$ (blue solid line) experiences collapses and revivals, a phenomenon systematically studied by J. H.

Eberly and his group with analytic approaches [52, 53]. The library gives the possibility to define any kind of operator, for example, the photon number parity operator, a very relevant operator for the Jaynes-Cummings model, which is defined as:

$$\Pi = e^{i\pi a^\dagger a}. \tag{34}$$

Its expectation value is also shown in Fig.10 (red solid line), and one observes that at the midpoint of the period of the collapse of the atomic inversion, where the field switches between states of even and odd parity, the parity expectation value is rapidly oscillating, a phenomenon coming from quantum interference effects [51].

## 8. Control in Quantum Technologies

The TorchQC library offers many useful tools and the ability to simulate and even control quite complex quantum systems that appear in modern quantum technologies. One such important system, studied in Ref. [54], is a quantum bus consisting of $N$ identical and non-interacting qubits coupled with a single-mode driven resonator and is closely related to the Dicke Model [55] of Quantum Optics. The system is described in the interaction picture by the following Hamiltonian:

$$H = \sum_{j=1}^{N} g_j(t) \left( a^\dagger \sigma_j^i + a\sigma_j^+ \right) + \xi(t)(a + a^\dagger), \tag{35}$$

where $g_j(t)$ denotes the time dependent coupling of the $j$-th qubit with the resonator and $\xi(t)$ is the time-dependent amplitude of the resonator drive. Following the methods proposed in our recent work [27], a deep learning model for the control functions can be easily implemented in the Pytorch library and the dynamics of the quantum system can be simulated with the TorchQC library. The proposed model with the name Black Box Neural Network (BBNN) is a fully connected feed-forward neural network inspired by physics-informed neural networks [56, 57, 58]. BBNN is a simpler implementation since it is employed only to represent the control functions and not the evolution of the quantum system [27]. The implementation of the method can be found in Appendix A.

The quantum control problem, studied in Ref. [54] with genetic optimization algorithms, is to efficiently drive this system to the following GHZ state:

$$|GHZ\rangle = \frac{|000\rangle + |111\rangle}{\sqrt{2}}. \tag{36}$$

The objective function the deep learning model minimizes is the infidelity of the final state with respect to the GHZ state:

$$\mathcal{F} = \langle GHZ|\psi(T)\rangle\,, \tag{37}$$

with $T$ being the final time of the quantum state evolution. At the end of the training phase, the neural network gives the control functions displayed in Fig. 11, while the corresponding evolution of the fidelity with the GHZ state is shown in Fig. 12.

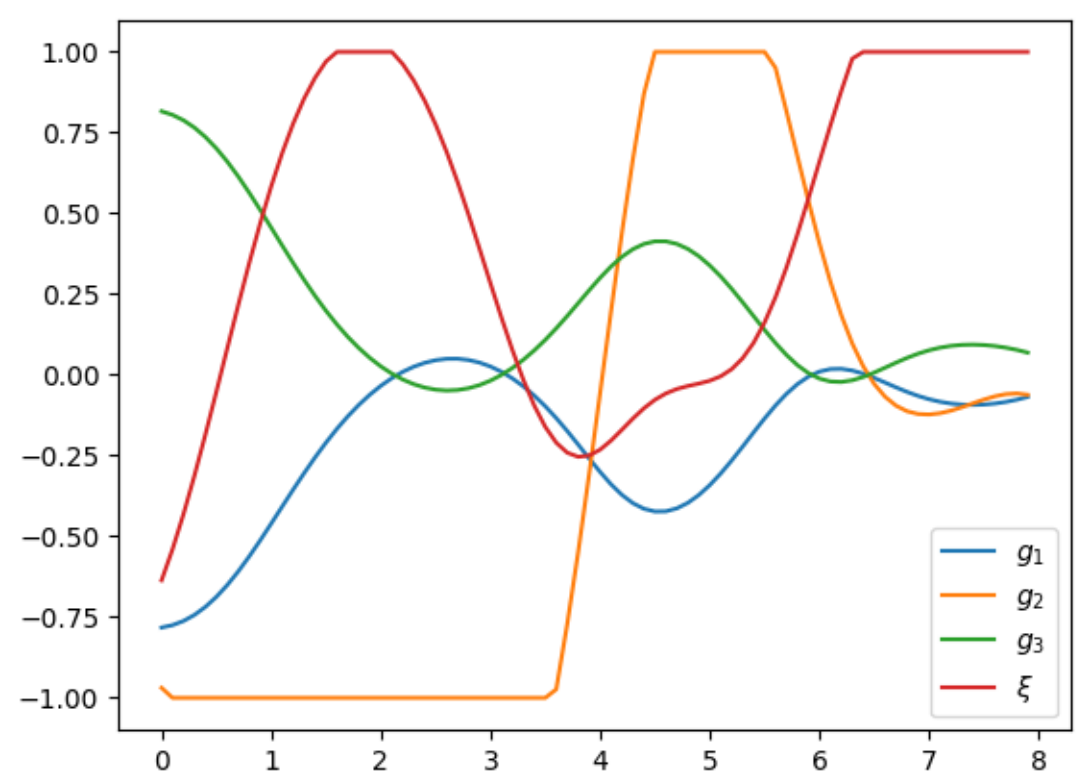

Figure 11: Control functions driving a three-qubit quantum bus, coupled with a single-mode resonator, to the GHZ state.

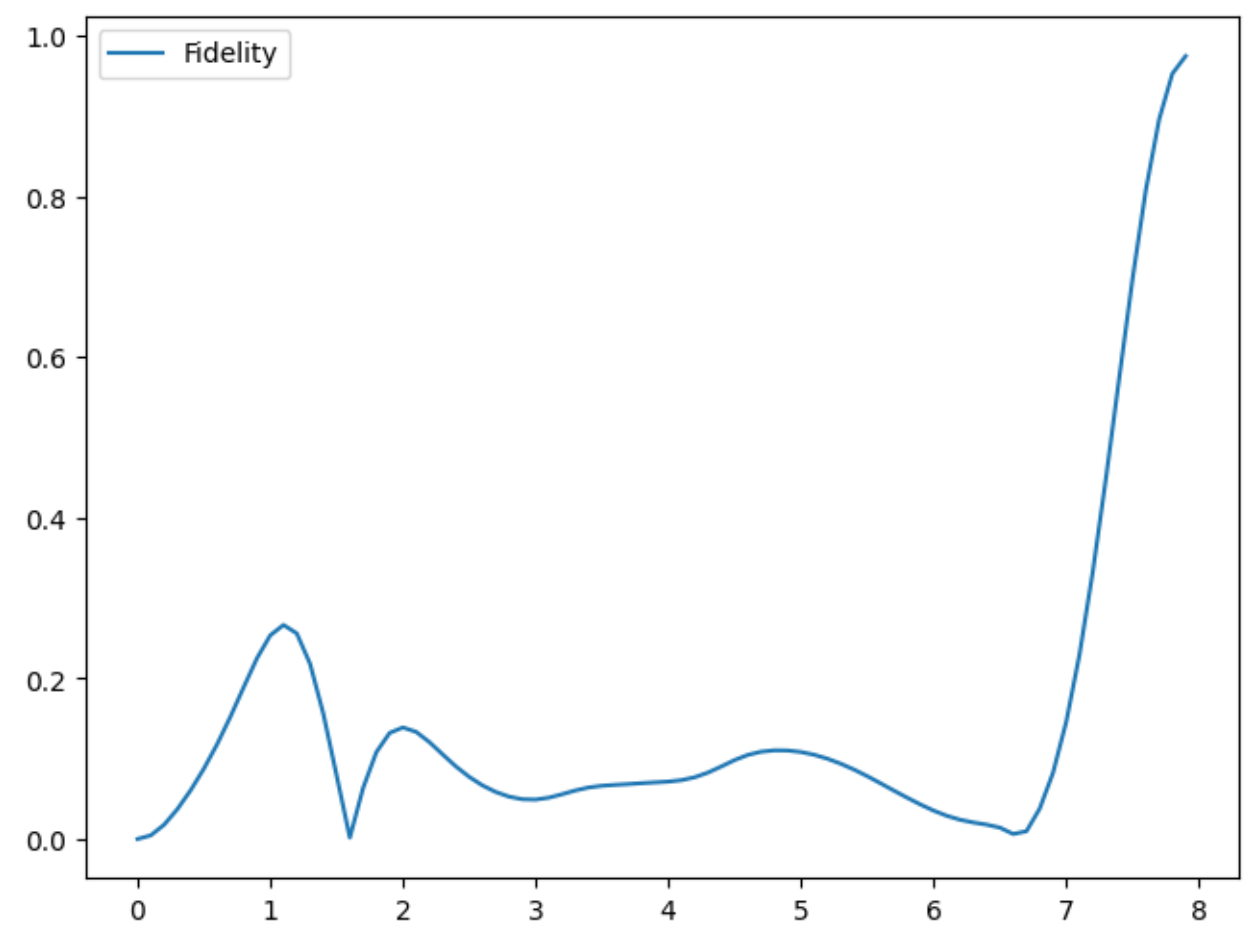

Figure 12: Corresponding evolution of the fidelity of the quantum bus state with respect to the GHZ state.

The final fidelity achieved is 0.9749, similar to that obtained with genetic algorithm methods in Ref. [54]. A higher fidelity is achieved if, following the authors' suggestion, an extra term is added to the loss function:

$$\phi = -\frac{0.1}{T}\int_0^T \langle n|\rho_{cav}(t)n\rangle \, dt, \tag{38}$$

punishing the large populations in higher Fock states $|n\rangle$ [54]. Training with this term leads to final fidelity equal to 0.9925.

## 9. Discussion

Previous sections demonstrate the power and applicability of the TorchQC library. It contains all the necessary structures to represent quantum states and quantum operators, as well as the methods needed to simulate the dynamics of both closed and open quantum systems. There are analytical methods such as the Time Dependent Schrödinger Equation (TDSE) and the Fock-Liouville space dynamics, as well as numerical methods with the von Neumann - Lindblad master equation and numerical TDSE. These form the basis where users can utilize ready methods to represent quantum systems and simulate quantum dynamics. This is, though, something that other packages such as QuTip and QuantumOptics.jl library can do quite effectively and with great variety and options. The main advantage of the suggested TorchQC library is its ability to easily integrate machine and deep learning models into the loop, making it easy to combine them with quantum dynamics, and the capacity to construct differentiable models directly by exploiting the automatic differentiation offered by the PyTorch library. Another advantage of the TorchQC library is that its routines can be executed in GPUs since it is an extension of PyTorch, which may potentially lead to substantial speedup of the simulations and the machine learning models that can be constructed.

As already demonstrated in Sections 6 and 8, using the suggested TorchQC library is quite straightforward to create a differentiable machine learning model with parameters for implementing algorithms such as GRAPE, as well as to utilize PyTorch to find optimal policies with reinforcement learning or to create deep learning models such as BBNN which can generate control sequences that effectively drive a quantum system to a target state. This work provides a way to easily introduce and incorporate machine learning models into quantum dynamics, and we thus believe that it will accelerate the integration of these powerful tools in quantum control for the ultimate goal to efficiently implement modern quantum technologies.

In terms of computational complexity, both analytical and numerical methods used for quantum dynamics scale linearly $\mathcal{O}(xn)$ with the number $(n)$ of time steps used to simulate the dynamics of the same quantum system, with $x$ being different for each method used. In terms of memory complexity, all methods scale linearly with the number of time steps $\mathcal{O}(xn)$, since the quantum states for all time steps are saved and returned by the corresponding methods at the end of the time evolution. With respect to the size of the quantum system, computational complexity follows the increase of the dimension of the associated Hilbert space. As a result, the complexity of many-body quantum simulations increases exponentially with the number of qubits. This curse of dimensionality poses restrictions on the scalability of classical simulations of large quantum systems and is common in all the frameworks used for this purpose. A potential advantage of the proposed framework is the possibility to run models, simulations and existing optimal control algorithms such as GRAPE in a powerful GPU, which could offer computational speedup compared to the conventional CPU execution. GPUs provide great parallelism and more efficient matrix multiplication and exponentiation, operations that both lie at the core of our simulation methods as well as in machine learning routines and algorithms like GRAPE. However, the main contribution of the present work is the possibility to easily create machine and deep learning models as alternative ways to study and control quantum systems, especially for multi-qubit systems where conventional optimal control methods may be difficult to apply. The next development which we intend to incorporate into our framework in the near future to improve the performance with respect to complexity are methods such as tensor networks [59, 60, 61, 62], which are capable of handling many-body quantum simulations more efficiently by sacrificing an unimportant amount of accuracy and can be also used for simulating non-Markovian open quantum system dynamics [60, 62].

### Acknowledgments

E.P. acknowledges support by the EU HORIZON-Project 101080085—QCFD.

## Appendix A.

### *Appendix A.1. Fock-Liouville Space dynamics*

The following code snippet demonstrates how one can simulate quantum dynamics in the Fock-Liouville space. The procedure is quite the same as in the Lindblad master equation method. The Fock-Liouville space is performed under the hood by the library and the result of the dynamics is the list of density matrices at the discrete time steps.

```
1 import numpy as np
2 import torch
3 import matplotlib.pyplot as plt
4
5 from torchqc.states import QuantumState
6 from torchqc.common_functions import get_density_matrix,\
7     expect_val_dm
```

```
8  from torchqc.operators import Operator, DynamicOperator
9  from torchqc.common_matrices import sigmaZ, sigmaX
10 from torchqc.dynamics import libland_equation
11 from torchqc.dynamics import libland_equation_FLS
12 
13 # define the initial condition
14 T = 5
15 Dt = 0.02
16 time = np.arange(0, T, Dt, dtype = np.float32)
17 D = 2 * np.pi
18 
19 # define hamiltonian
20 T = 10
21 Dt = 0.1
22 time = np.arange(0, T + Dt, Dt, dtype = np.float32)
23 
24 H = sigmaX()
25 hamiltonian = DynamicOperator(2, H, time=time)
26 
27 # simulate dynamics with libland equation
28 rho = get_density_matrix(state)
29 _, states = libland_equation(rho, hamiltonian,
30     time, Dt, [sigmaX()], [0.01 * 2 * np.pi])
31 
32 P1 = np.array(
33     [torch.real(state.matrix[0][0]) for state in states]
34     )
35 P2 = np.array(
36     [torch.real(state.matrix[1][1]) for state in states]
37     )
38 expect_Z_list = expect_val_dm(states, sigmaZ())
39 
40 # simulate dynamics in the Fock-Liouville space
41 _, states = libland_equation_FLS(rho, hamiltonian,
42     time, Dt, [sigmaX()], [0.01 * 2 * np.pi])
43 expect_Z_list_FLS = expect_val_dm(states, sigmaZ())
44 P1_FLS = np.array(
45     [torch.real(state.matrix[0][0]) for state in states]
46 )
47 P2_FLS = np.array(
48     [torch.real(state.matrix[1][1]) for state in states]
```

```
)

# plot results for comparizon
fig, ax = plt.subplots()

ax.plot(time, P1, label = "P1")
ax.plot(time, P2, label = "P2")
ax.scatter(time, P1_FLS, marker='x', label = "P1 FLS")
ax.scatter(time, P2_FLS, marker='x', label = "P2 FLS")
ax.legend()
fig.show()

fig, ax = plt.subplots()

ax.plot(time, expect_Z_list, label = r"$<\sigma_Y>$")
ax.scatter(time, expect_Z_list_FLS,
    marker='x', label = r"$<\sigma_Y> FLS$")
ax.set_xlabel("t")
ax.set_xlabel("t")
ax.legend()
fig.show()
```

Listing 9: Fock-Liouville space dynamics code

### *Appendix A.2. GRAPE implementation with Automatic Differentiation*

The code that implements the GRAPE algorithm for the population inversion problem in a qubit with the help of the TorchQC library is presented bellow.

```
import numpy as np
import matplotlib.pyplot as plt
import torch
import torch.nn as nn

from torchqc.states import QuantumState
from torchqc.common_matrices import sigmaX, sigmaZ, \
    sigmaY

torch.manual_seed(123987456)

# define the Unitary gate as a module
```

```python
# of the library Pytorch
max_control = 1.

class UnitatyGate(nn.Module):
    def __init__(self, nb_params = 1, state_size = 3,
                 ham_terms = [
                     sigmaX(), sigmaZ(), sigmaY()],
                     Dt = 0.1):
        super(UnitatyGate, self).__init__()

        self.nb_params = nb_params
        self.ham_terms = ham_terms

        self.params = nn.Parameter(
            torch.randn(nb_params))
        self.Dt = Dt

    def forward(self, input_state):
        Ht = torch.zeros_like(
            self.ham_terms[0].matrix)

        for i, operator in enumerate(self.ham_terms):
            Ht += torch.clip(
                self.params[i], -max_control,
                max_control
                ) * operator.matrix

        t0 = 0
        t1 = self.Dt
        time = np.linspace(t0, t1, 2)
        hamiltonian = DynamicOperator(2,
            Operator(2, Ht), time=time)

        states = TDSE(QuantumState(2, input_state),
            hamiltonian, time=time, Dt=self.Dt)

        output_state = states[-1].state_tensor

        return output_state

class UnitariesProduct(nn.Module):
```

```
    def __init__(self, nb_unitaries, nb_unitaty_params,
                state_size, ham_terms=[
                    sigmaX(), sigmaZ()],
                    Dt = 0.1):
        super(UnitariesProduct, self).__init__()

        self.nb_unitaries = nb_unitaries
        self.unitaries = []
        self.params = []
        self.states = torch.zeros(tuple([nb_unitaries]) \
                    + state_size, dtype=torch.complex128)

        for i in range(nb_unitaries):
            new_unitaty = UnitatyGate(nb_unitaty_params,
                state_size, ham_terms, Dt)
            self.unitaries.append(new_unitaty)

            self.register_parameter(f"params{i}",
                new_unitaty.params)

    def forward(self, input_state):
        out = input_state

        for i, unitary in enumerate(self.unitaries):
            out = unitary(out)
            self.states[i] = out

        return out

# system parameters
T = 3.15
nb_unitaries = 100
Dt = T / nb_unitaries

initial_state = QuantumState.basis(2)[0]
target_state = QuantumState.basis(2)[1]

# initialize model
model = UnitariesProduct(nb_unitaries=nb_unitaries,
    nb_unitaty_params=2,
    state_size=tuple(initial_state.state_tensor.shape),
```

```
95      Dt=Dt)
96
97  def fidelity(output_state, target_state):
98      return torch.abs(
99          torch.inner(
100         output_state.view(target_state.dims),
101         target_state.state_tensor.view(target_state.dims))
102         ) ** 2
103
104 def loss_fn(output_state, states, add_params=True):
105     loss = 0
106     target_loss = 1 - fidelity(output_state,
107         target_state)
108
109     loss += target_loss
110
111     if add_params:
112         weight = 0.001
113
114         for param in model.parameters():
115             loss += weight * torch.norm(param) ** 2
116
117     return loss, target_loss
118
119 # define training function
120 def training():
121     losses = []
122     loss_threshold = 1e-4
123     iterations = 2000
124
125     model.train()
126
127     learning_rate = 1e-3
128     optimizer = torch.optim.Adam(model.parameters(),
129         lr=learning_rate)
130     print_interval = 100
131     num_epochs = 0
132
133     for iter in range(iterations):
134         # forward pass
135         output_state = model(initial_state.state_tensor)
```

```
136
137         # calculate loss based on controls
138         # that produced by the nn
139         loss, target_loss = loss_fn(
140             output_state,
141             model.states
142         )
143
144         losses.append(loss.detach().numpy())
145
146         with torch.no_grad():
147             infidelity = target_loss
148
149             if num_epochs % print_interval == 0:
150                 print("Epoch = ", num_epochs, ",
151                     Loss = ",
152                     infidelity.detach().numpy())
153
154             if infidelity < loss_threshold:
155                 break
156
157         # backpropagation
158         optimizer.zero_grad()
159         loss.backward()
160         optimizer.step()
161
162         num_epochs += 1
163
164 # start training
165 training()
166
167 # get results
168 with torch.no_grad():
169     output_state = model(initial_state.state_tensor)
170     _, infidelity = loss_fn(output_state,
171         model.states)
172     print("Fidelity = ", 1 - infidelity.numpy())
173
174     omegas = []
175     detunings = []
176
```

```
    for param in model.parameters():
        param = torch.clip(param, -max_control,
            max_control)

        omega = param[0].numpy()
        delta = param[1].numpy()

        omegas.append(omega)
        detunings.append(delta)

    states = model.states

    population1 = [
        torch.abs(state[0]) ** 2 for state in states
    ]
    population2 = [
        torch.abs(state[1]) ** 2 for state in states
    ]

# plot unitary propagator control values
time = np.arange(0, T, Dt, dtype = np.float32)
fig, ax = plt.subplots(2, 1)
ax1 = ax[0]
ax2 = ax[1]

ax1.step(time, omegas, label = r"$\Omega$")
ax2.step(time, detunings, color = 'g', label = r"$\Delta$")
ax1.legend()
ax2.legend()

ax1.set_title('(a)', loc = "right", fontsize = 10)
ax2.set_title('(b)', loc = "right", fontsize = 10)
fig.show()

# plot populations evolution
fig, ax = plt.subplots()
ax.plot(time, population1, label = "P1")
ax.plot(time, population2, label = "P2")
ax.legend()
fig.show()
```

Listing 10: GRAPE code

*Appendix A.3. MDP environment for the qubit system*

The following snippet creates the qubit environment as a Markov Decision Process (MDP), by defining the reset, step and specification creation functions. The reset function brings the environment state to the initial position, the step function defines the way the action of the agent should act on the state of the environment, and the specification creation function defines how the state and actions spaces should be defined for the qubit system case.

```
import numpy as np

import matplotlib.pyplot as plt
import torch
from tensordict import TensorDict, TensorDictBase
from torchrl.data import BoundedTensorSpec, \
    CompositeSpec, UnboundedContinuousTensorSpec
from torchrl.envs import EnvBase
from torchrl.envs.utils import check_env_specs, step_mdp
from typing import Optional

from qctorch.operators import Operator
from qctorch.states import QuantumState
from qctorch.common_functions import get_density_matrix,\
    fidelity
from qctorch.dynamics import libland_equation
from qctorch.common_matrices import sigmaX

## DEFINE THE FUNCTIONS
def make_composite_from_td(td):
    # custom function to convert a ``tensordict`` in
    # a similar spec structure of unbounded values.
    composite = CompositeSpec(
        {
            key: make_composite_from_td(tensor)
            if isinstance(tensor, TensorDictBase)
            else UnboundedContinuousTensorSpec(
                dtype=tensor.dtype, device=tensor.device,
                    shape=tensor.shape
            )
            for key, tensor in td.items()
        },
        shape=td.shape,
```

```
34      )
35      return composite
36
37  def _make_spec(self, td_params):
38      dims = td_params["params", "dims"].numpy()
39
40      # define the observation / state space specs
41      observation_shape = 8 # density matrix elements
42
43      self.observation_spec = CompositeSpec(
44          observation=BoundedTensorSpec(
45              low=torch.zeros(observation_shape,
46                  dtype=torch.float32),
47              high=torch.ones(observation_shape,
48                  dtype=torch.float32),
49              shape=observation_shape,
50              dtype=torch.float32
51          ),
52          params=make_composite_from_td(td_params["params"]),
53          shape=(),
54      )
55
56      self.state_spec = self.observation_spec.clone()
57
58      # define the action space specifications
59      self.action_spec = BoundedTensorSpec(
60          low=torch.tensor([-1]),
61          high=torch.tensor([1]),
62          shape=(1,),
63          dtype=torch.float32
64      )
65
66      # define reward spec
67      self.reward_spec = BoundedTensorSpec(
68          low=torch.tensor([-1]),
69          high=torch.tensor([1]),
70          shape=(1,),
71          dtype=torch.float32
72      )
73
74  def _step(self, tensordict):
```

```
    # get the initial density matrix for this time step
    # initial_dm = tensordict["density_matrix"]
    initial_dm = self.curr_state

    # get the action, which is the Rabi frequency
    actions_tensor = tensordict["action"]
    actions = actions_tensor.numpy()
    [omega] = actions

    # dt = tensordict["params", "dt"].numpy()
    dt = 0.04
    target_idx = tensordict["params", "target_state"]
    target_idx = target_idx.numpy()
    dims = tensordict["params", "dims"].numpy()
    max_steps = tensordict["params", "max_steps"].numpy()

    t0 = 0
    T = dt
    time = np.linspace(t0, T, 2)

    self.current_step += 1

    hamiltonian = Operator(2,
        (omega / 2) * sigmaX().matrix.expand(
            len(time), -1, -1)
        )

    _, states = libland_equation(initial_dm, hamiltonian,
        time, dt)

    final_dm = states[-1]

    self.curr_state = final_dm

    final_dm_as_real = torch.
        view_as_real(final_dm.matrix)
    final_dm_as_real = final_dm_as_real.
        type(torch.float32).flatten()

    target_state = get_density_matrix(
        QuantumState.basis(dims)[target_idx])
```

```
116
117     fidelity = torch.real(final_dm.matrix[1][1])
118
119     reward = fidelity - 1 # punish each step by -1
120
121     # it should be true on time T
122     # or when fidelity achieved
123     if fidelity > 0.999 or self.current_step >= max_steps:
124         if fidelity >= 0.999:
125             reward += 10
126         elif fidelity >= 0.9999:
127             reward += 10
128
129         done = torch.ones_like(reward, dtype=torch.bool)
130     else:
131         done = torch.zeros_like(reward, dtype=torch.bool)
132
133     # create and return new tensordict
134     out = TensorDict(
135         {
136             "observation": final_dm_as_real,
137             "density_matrix": final_dm,
138             "params": tensordict["params"],
139             "reward": reward,
140             "done": done,
141         },
142         tensordict.shape,
143     )
144     return out
145
146 def _reset(self, tensordict):
147     if tensordict is None or tensordict.is_empty():
148         tensordict = QubitEnv.gen_params()
149
150     self.current_step = 0
151
152     # define initial state with params
153     state_idx = tensordict["params", "initial_state"]
154         .numpy()
155     dims = tensordict["params", "dims"].numpy()
156     initial_state = QuantumState.basis(dims)[state_idx]
```

```python

    # get initial density matrix
    rho = get_density_matrix(initial_state)

    self.curr_state = rho

    rho_as_real = torch.view_as_real(rho.matrix)
    rho_as_real = rho_as_real.type(torch.float32).flatten()

    # create out tensordict
    out = TensorDict(
        {
            "observation": rho_as_real,
            "density_matrix": rho,
            "params": tensordict["params"],
        },
        batch_size=tensordict.shape,
    )

    return out

def _set_seed(self, seed: Optional[int]):
    rng = torch.manual_seed(seed)
    self.rng = rng

def gen_params(batch_size=None) -> TensorDictBase:
    """Returns a ``tensordict``
    containing the physical parameters"""

    if batch_size is None:
        batch_size = []
    td = TensorDict(
        {
            "params": TensorDict(
                {
                    "max_control": 1.,
                    "dims": 2,
                    "dt": 0.04,
                    "T": 4,
                    "initial_state": 0,
                    "target_state": 1,
```

```
198                     "max_steps": 100
199                 },
200                 [],
201             )
202         },
203         [],
204     )
205 
206     if batch_size:
207         td = td.expand(batch_size).contiguous()
208     return td
209 
210 
211 ## DEFINE THE ENVIRONMENT CLASS
212 class QubitEnv(EnvBase):
213     metadata = {}
214     batch_locked = True
215     current_step = 0
216 
217     def __init__(self, td_params=None, seed=None,
218         device="cpu"):
219         if td_params is None:
220             td_params = self.gen_params()
221 
222         super().__init__(device=device, batch_size=[])
223         self._make_spec(td_params)
224 
225         if seed is None:
226             seed = torch.empty((),
227                 dtype=torch.int64).random_().item()
228         self.set_seed(seed)
229 
230     gen_params = staticmethod(gen_params)
231     _make_spec = _make_spec
232 
233     # Mandatory methods: _step, _reset and _set_seed
234     _reset = _reset
235     _step = _step
236     _set_seed = _set_seed
```

Listing 11: Qubit MDP environment code

### *Appendix A.4. Jaynes-Cumming model dynamics*

Jaynes-Cummings Hamiltonian (32) can be easily constructed in the TorchQC library and, using the already implemented method of Lindblad master equation, the model can be simulated. Lines 34 - 46 in the following code snippet show how to implement this Hamiltonian which, since it will be used for dynamics simulations, it is easier and better to be defined as a DynamicOperator class, as already described in section 3.

```
1  import numpy as np
2  import matplotlib.pyplot as plt
3
4  from torchqc.states import QuantumState
5  from torchqc.common_functions import get_density_matrix,\
6      expect_val_dm
7  from torchqc.operators import DynamicOperator
8  from torchqc.common_matrices import sigmaZ, annihilation,\
9      creation, eye, sigmaPlus, sigmaMinus
10 from torchqc.dynamics import libland_equation
11 from torchqc.tensor_product import tensor_product_ops,\
12     tensor_product_states
13
14 #Define system parameters and initial coherent state
15 n = 2 # two level atom
16 N = 50 # number of fock states
17 T = 40
18 Dt = 0.1
19 wc = 0.1  # cavity frequency
20 wa = 0.1  # atom frequency
21 g = 1  # coupling strength
22 time = np.arange(0, T, Dt, dtype = np.float32)
23
24 total_dims = n * N
25 basis_states = QuantumState.basis(n)
26 atom_state = basis_states[0]
27 b = np.sqrt(20)
28 coherent_state = QuantumState.coherent(N, b)
29
30 initial_state = tensor_product_states(coherent_state,
31     atom_state)
32
33 # Define Jaynes-Cummings Hamiltonian under RWA
```

```
a_dagger = creation(N)
a = annihilation(N)
sigmap = sigmaPlus()
sigmam = sigmaMinus()

H1 = tensor_product_ops(eye(N), -(wa / 2) * sigmaZ())
H2 = tensor_product_ops(wc * a_dagger * a, eye(n))
H3 = g * (tensor_product_ops(a, sigmap) +
    tensor_product_ops(a_dagger, sigmam))
H = H1 + H2 + H3

hamiltonian = DynamicOperator(dims=total_dims,
    Ht=H, time=time)

# Simulate dynamics
rho = get_density_matrix(initial_state)
time_tensor, states = libland_equation(rho,
    hamiltonian, time, Dt)

# simulate dynamics with Libland-von-Neumann equation
N_op = a_dagger * a
expectation_values = expect_val_dm(states,
    tensor_product_ops(eye(N), sigmaZ()))
expectation_values2 = expect_val_dm(states,
    tensor_product_ops(N_op, eye(n)))

# plot results
fig, ax = plt.subplots()

plt.plot(time, expectation_values,
    label=r"W(t) = $<\sigma_z>$")
plt.xlabel("gt")
plt.ylabel("W(t)")
plt.title("Oscillations collapse and revival")
ax.legend()
fig.show()
```

Listing 12: Jaynes-Cummings model dynamics code

*Appendix A.5. Black Box Neural Network (BBNN) combined with TorchQC*

The following snippet demonstrates how one can implement a deep learning model, specifically a feed-forward neural network, and how this can be used to approximate the control functions. By training the model, the controls that it produces become more and more optimal as it tries to minimize the objective function.

```
framesep=2mm,
baselinestretch=1.2,
bgcolor=LightGray,
fontsize=\footnotesize,
linenos]{python}
import numpy as np
import torch
import torch.nn as nn
import matplotlib.pyplot as plt

from torchqc.states import QuantumState
from torchqc.common_functions import get_density_matrix,\
    expect_val_dm, fidelity
from torchqc.operators import DynamicOperator
from torchqc.common_matrices import sigmaZ, annihilation,\
    creation, eye, sigmaMinus, sigmaPlus
from torchqc.dynamics import TDSE
from torchqc.tensor_product import tensor_product_ops,\
    tensor_product_states, partial_trace

torch.manual_seed(1989)

# define the neural network
max_control = 1.

class NeuralNet(nn.Module):
    def __init__(self, hidden_size, nb_hidden_layers = 2,
            input_size = 1, output_size = 1,
            activation_fn = nn.Tanh):
        super(NeuralNet, self).__init__()

        self.nb_hidden_layers = nb_hidden_layers
        self.hidden_layers = []
        self.hidden_act_layers = []
```

```
35
36         # input layer
37         self.input_layer = nn.Linear(input_size,
38             hidden_size)
39         self.relu_input = activation_fn()
40
41         # hidden layers
42         for layer in range(nb_hidden_layers):
43             new_layer = nn.Linear(hidden_size,
44                 hidden_size)
45             self.hidden_layers.append(new_layer)
46             self.hidden_act_layers.append(
47                 activation_fn()
48             )
49
50             # hidden parameters should be registered
51             self.register_parameter(f"weights_{layer}",
52                 new_layer.weight)
53             self.register_parameter(f"bias_{layer}",
54                 new_layer.bias)
55
56         # output layer
57         self.output_layer = nn.Linear(hidden_size,
58             output_size)
59
60     def forward(self, x):
61         out = self.input_layer(x)
62         out = self.relu_input(out)
63
64         for layer in range(self.nb_hidden_layers):
65             out = self.hidden_layers[layer](out)
66             out = self.hidden_act_layers[layer](out)
67
68         out = self.output_layer(out)
69         out = torch.clip(input=out, min=-max_control,
70             max=max_control)
71
72         return out
73
74 class Sin(nn.Module):
75     """The sin activation function.
```

```python
    """

    def __init__(self):
        """Initializer method.
        """
        super().__init__()

    def forward(self, input_):
        return torch.sin(input_)

# Initial qubits state |010>
#  with resonator in ground state |0>
N = 10 # resonator modes
n = 2
# tensor product of two qubits and a resonator mode
total_dims = n * n * n * N

qubit_states_basis = QuantumState.basis(n)
resonator_basis_states = QuantumState.basis(N)

qubit1_state = qubit_states_basis[0]
qubit2_state = qubit_states_basis[1]
qubit3_state = qubit_states_basis[0]
resonator_state = resonator_basis_states[0]

initial_state = tensor_product_states(
    qubit1_state,
    qubit2_state,
    qubit3_state,
    resonator_state)

target_state = (1 / np.sqrt(2)) * (
        tensor_product_states(qubit_states_basis[0],
            qubit_states_basis[0], qubit_states_basis[0]
    ) + tensor_product_states(
        qubit_states_basis[1],
        qubit_states_basis[1],
        qubit_states_basis[1])
    )

target_state = get_density_matrix(target_state)
```

```

# simulate dynamics with TDSE
tau = 1
T = 8 * tau
Dt = 0.1 * tau
time = np.arange(0, T, Dt, dtype = np.float32)
time_tensor = torch.from_numpy(time)
    .reshape(len(time), 1)

# Initialize the BBNN
BBNN = NeuralNet(hidden_size=150, input_size=1,
    output_size=4, nb_hidden_layers=4, activation_fn=Sin)

# define quantum simulation and loss functions
def quantum_simulation(u_pred):
    a_dagger = creation(N)
    a = annihilation(N)
    sm = sigmaMinus()
    sp = sigmaPlus()

    g1 = u_pred[:, 0:1]
    g1.unsqueeze_(-1)
    g1.expand(len(time), total_dims, total_dims)

    g2 = u_pred[:, 1:2]
    g2.unsqueeze_(-1)
    g2.expand(len(time), total_dims, total_dims)

    g3 = u_pred[:, 2:3]
    g3.unsqueeze_(-1)
    g3.expand(len(time), total_dims, total_dims)

    xi = u_pred[:, 3:4]
    xi.unsqueeze_(-1)
    xi.expand(len(time), total_dims, total_dims)

    H1 = g1 * (tensor_product_ops(sm, eye(n),
        eye(n), a_dagger) +\
        tensor_product_ops(sp, eye(n), eye(n), a)) \
    + g2 * (tensor_product_ops(eye(n), sm,
        eye(n), a_dagger) \
```

```
158     + tensor_product_ops(eye(n), sp, eye(n), a)) \
159         + g3 * (tensor_product_ops(eye(n),
160             eye(n), sm, a_dagger)\
161         + tensor_product_ops(eye(n), eye(n), sp, a))
162 
163     H2 = xi * tensor_product_ops(eye(n), eye(n),
164         eye(n), a_dagger + a)
165 
166     H = H1 + H2
167 
168     hamiltonian = DynamicOperator(total_dims, H.matrix)
169     states = TDSE(initial_state, hamiltonian, time, Dt)
170 
171     return states
172 
173 def criterion_fidelity_custom(u_pred):
174     states = quantum_simulation(u_pred)
175 
176     achieved_state = states[-1]
177     achieved_rho = get_density_matrix(achieved_state)
178     reduced_achieved_state = partial_trace(achieved_rho,
179         [3])
180 
181     infidelity = 1 - fidelity(reduced_achieved_state,
182         target_state)
183 
184     return infidelity
185 
186 # define training function and start training
187 def training(learning_rate=1e-3, iterations=100):
188     losses = []
189     loss_threshold = 1e-4
190 
191     BBNN.train()
192 
193     optimizer = torch.optim.Adam(BBNN.parameters(),
194         lr=learning_rate)
195     print_interval = 10
196     num_epochs = 0
197 
198     # while loss_float >= loss_threshold:
```

```
199     for iter in range(iterations):
200         # forward pass
201         u_pred = BBNN(time_tensor)
202 
203         # calculate loss based on controls
204         # that produced by the nn
205         loss = criterion_fidelity_custom(u_pred)
206         losses.append(loss.detach().numpy())
207 
208         # backpropagation
209         optimizer.zero_grad()
210         loss.backward()
211         optimizer.step()
212 
213         if num_epochs % print_interval == 0:
214             print("Epoch = ", num_epochs,
215                 ", Infidelity = ",
216                 loss.clone().detach().numpy())
217 
218         num_epochs += 1
219 
220         if losses[-1] < loss_threshold:
221             break
222 
223 # Two training sessions with different learning rates
224 training(1e-3, 500)
225 training(1e-4, 500)
226 
227 # Get the optimal control functions values
228 with torch.no_grad():
229     u_pred = BBNN(time_tensor)
230 
231     states = quantum_simulation(u_pred)
232 
233     g1 = u_pred[:, 0:1].detach().numpy()
234     g1 = np.array([omega[0] for omega in g1])
235 
236     g2 = u_pred[:, 1:2].detach().numpy()
237     g2 = np.array([omega[0] for omega in g2])
238 
239     g3 = u_pred[:, 2:3].detach().numpy()
```

```
240     g3 = np.array([omega[0] for omega in g3])
241
242     xi = u_pred[:, 3:4].detach().numpy()
243     xi = np.array([omega[0] for omega in xi])
244
245 # Print results
246 fig, ax = plt.subplots()
247
248 ax.plot(time, g1, label = r"$g_1$")
249 ax.plot(time, g2, label = r"$g_2$")
250 ax.plot(time, g3, label = r"$g_3$")
251 ax.plot(time, xi, label = r"$\xi$")
252 ax.set_ylim(-1.1, 1.1)
253 ax.legend()
254 ax.show()
255
256 # compute and print fidelities
257 fidelities = torch.tensor(
258     [fidelity(partial_trace(get_density_matrix(state),
259         [3]), target_state) for state in states]
260     )
261
262 fig, ax = plt.subplots(layout='constrained')
263
264 ax.plot(time, fidelities, label=r"Fidelity")
265 ax.legend()
266 ax.show()
```

Listing 13: BBNN code

## Listings